\documentclass[conference]{IEEEtran}

\usepackage{cite}
\usepackage{hyperref}
\usepackage{graphicx}

\usepackage[caption=false]{subfig}

\usepackage{multicol}
\usepackage{listings}
\usepackage{color}
\usepackage[numbers]{natbib}

\definecolor{darkblue} {rgb}{0.0, 0.0, 0.8}
\definecolor{darkgreen}{rgb}{0.0, 0.4, 0.0}
\definecolor{mauve}    {rgb}{0.6, 0.0, 0.5}

\newcommand{\fulltitle}{Myrmics: Scalable, Dependency-aware Task
                         Scheduling on Heterogeneous Manycores}

\begin{document}


\title{\fulltitle}
\author{
  \IEEEauthorblockN{Spyros Lyberis, Polyvios Pratikakis, Iakovos Mavroidis}
  \IEEEauthorblockA{
    Institute of Computer Science,\\
    Foundation for Research \& Technology - Hellas
  }
  \and
  \IEEEauthorblockN{Dimitrios S. Nikolopoulos}
  \IEEEauthorblockA{
    School of Electronics, Electrical Engineering\\
    and Computer Science\\
    Queen's University of Belfast
  }
}

%
%
%


\maketitle

\begin{abstract}


Task-based programming models have become very popular, as they
offer an attractive solution to parallelize serial application
code with task and data annotations. They usually depend on a
runtime system that schedules the tasks to multiple cores in
parallel while resolving any data hazards. However, existing
runtime system implementations are not ready to scale well on
emerging manycore processors, as they often rely on centralized
structures and/or locks on shared structures in a cache-coherent
memory. We propose design choices, policies and mechanisms to
enhance runtime system scalability for single-chip processors
with hundreds of cores. Based on these concepts, we create and
evaluate \textit{Myrmics}, a runtime system for a
dependency-aware, task-based programming model on a heterogeneous
hardware prototype platform that emulates a single-chip processor
of 8 latency-optimized and 512 throughput-optimized CPUs. We find
that Myrmics scales successfully to hundreds of cores. Compared
to MPI versions of the same benchmarks with hand-tuned message
passing, Myrmics achieves similar scalability with a 10-30\%
performance overhead, but with less programming effort. 
We analyze the scalability of the runtime system in detail and
identify the key factors that contribute to it.

\end{abstract}

\section{Introduction}
\label{sec:intro}

Task-based programming models allow developers to express parallelism
at a higher level than threads.  A task-parallel program consists of
\textit{tasks}, which are relatively small function calls performing
atomic chunks of work that run to completion without communication
with other tasks.  A runtime system then schedules and executes the
tasks in parallel.
Tasks are usually annotated using compiler pragmas
and the resulting parallel program is an equivalent
implementation of the sequential one.  Without further assistance
from the programmer, the runtime system considers all spawned
tasks to be independent and eligible for execution. The OpenMP
support for tasks~\cite{openmp_tasks} falls into this category.
Many researchers advocate that if the programmer also provides
information on what data the task will operate, the runtime can
extract automatically the maximum amount of parallelism from the
program, reduce synchronization costs and guarantee determinism.
Examples on such programming models include Legion~\cite{legion},
Dynamic Out-of-Order Java~\cite{doj}, OmpSs~\cite{ompss} and
Data-Driven Tasks~\cite{data_driven_tasks}. Advantages of these
\textit{dependency-aware} models include increased
programmability, improved data locality and flexible exploitation
of parallelism depending on the application phase. 

Despite the increasing popularity of dependency-aware, task-based
programming models, there is much room for improvement. First,
many of these models do not support nested parallelism ---they
assume that a single master procedure spawns all tasks. As
processors scale to tens of CPU cores, the single master becomes
a bottleneck. Authors tend to evaluate their work running on at
most 64 cores. It remains largely unexplored how these runtime
systems will behave on processors with hundreds of cores. Second,
there is limited support for irregular, pointer-based data
structures, such as trees and graphs. In most models it is not
possible to parallelize accesses on such structures. Bauer et
al.~\cite{legion} have recently presented such concepts in
Legion. Concurrently with their work, we have introduced a
programming model~\cite{progr_model_mspc11} and a memory
allocator~\cite{myrmics_ismm12} to support irregular parallelism
using \textit{regions}, an efficient way to express arbitrary
collection of heap objects.  Third, existing runtimes do not
project well to future architectures.  Recent research from
Intel~\cite{runnemede,last_millennium} that explores CPU
architectures for years 2018 and beyond, argues for
specialized CPU cores for runtime and application code as well
as non-coherent caches, in order to increase energy efficiency.
They further claim that dataflow task-based models will be
particularly suited for such manycore chips. To the best of our
knowledge, most of the existing programming models do not take into
account these CPU architecture predictions. In particular, they
do not dedicate processor cores to runtime functions (and
specifically they do not consider heterogeneous single-chip
architectures with few strong cores and many weaker ones), they
do not consider software-assisted scalable memory coherence and
they are not based on scalable dependency analysis and scheduling
algorithms.

We introduce \textit{Myrmics}\footnote{\ The Myrmics source code,
documentation, examples, and benchmarks, are available at
\url{www.myrmics.com}.}, a runtime system that address these
shortcomings. Our work makes the following contributions:

\begin{itemize}

  \item We introduce hierarchical dependency analysis and
  scheduling algorithms for task dataflow models implemented on
  non cache-coherent, manycore architectures. We show
  experimentally that this enables scaling to hundreds of cores,
  alleviates the single-master and centralized structures
  bottlenecks and effectively trades locality with parallelism.

  \item We use hierarchical regions to efficiently support
  nested parallelism of pointer-based data structures. We
  integrate them in the hierarchical dependency analysis and
  scheduling algorithms to make them scalable. In previous
  work~\cite{progr_model_mspc11,myrmics_ismm12} we have provided
  a theoretical proof for the region-based programming model
  determinism and proposed region-based, scalable,
  memory management algorithms.

  \item Based on the aforementioned algorithms, we design
  Myrmics, a scalable task runtime system, to run on a
  heterogeneous, 520-core, non-coherent, prototype processor.
  Myrmics uses CPUs with different capabilities to run runtime
  and application code. Our system addresses the challenges a
  runtime will face on future processors, according to the
  current design trends.

  \item We evaluate Myrmics using several synthetic benchmarks, five
  standard kernels, and one irregular application. We analyze the
  trade-offs and overheads and we compare the results with reference
  MPI implementations on the same platform. Myrmics achieves similar
  scalability to the hand-tuned MPI implementations with a performance
  overhead between 10-30\%, but with less programming effort.

\end{itemize}

The rest of this paper is organized as follows.
Section~\ref{sec:model} overviews the task-based programming
model we use. Section~\ref{sec:platform} overviews the 520-core
hardware prototype we run Myrmics on. In
section~\ref{sec:choices} we discuss our key design choices for
the runtime system. Section~\ref{sec:design} presents the
Myrmics design and its subsystems implementation in greater
detail. In section~\ref{sec:evaluation} we evaluate the Myrmics
performance and overheads and pin it against reference MPI runs.
Finally, section~\ref{sec:related} presents related work and
in section~\ref{sec:discussion} we share our insights on
designing and evaluating Myrmics.

\section{Background: Programming Model}
\label{sec:model}

\begin{figure}
  \begin{lstlisting}
typedef struct {
  rid_t   lreg, rreg;
  struct  TreeNode *left, *right;
} TreeNode;
main() {
  rid_t     top;    // Whole tree@\label{src:top}@
  TreeNode *root; // Tree root@\label{src:root}@
  // ...
#pragma myrmics region inout(top)@\label{src:pragma1}@
    process(root);@\label{src:pragma1e}@
#pragma myrmics region in(top)@\label{src:pragma2}@
    print(root); @\label{src:pragma2e}@
}
void process(TreeNode *n) {
  if(n->left)
#pragma myrmics region inout(n->lreg) @\label{src:pragma3}@
    process(n->left);
  if(n->right)
#pragma myrmics region inout(n->rreg)
    process(n->right); @\label{src:pragma3e}@
}
void print(TreeNode *root) {
  print(root->left); @\label{src:print}@
  print_result(root);
  print(root->right); @\label{src:printe}@
}
  \end{lstlisting}
  \caption{Myrmics code example to hierarchically process, and
  then print, a binary tree.}
  \label{fig:code_example}
\end{figure}

We design the decentralized Myrmics runtime to implement a
dependency-aware, task-based programming model with region-based
memory management. Myrmics regions are dynamic, \emph{i.e.},
growable pools of memory that contain objects or subregions. In
past work~\cite{progr_model_mspc11}, we defined the semantics of
the programming model and formally proved that its parallel
execution on non-coherent systems is always deterministic and
equivalent to a serial execution. We use a source-to-source
compiler~\cite{scoop_pact12} to translate pragma-annotated C code
to plain C code with calls to the Myrmics API, which is described
in section~\ref{sec:design} (Fig.~\ref{fig:api}).

Fig.~\ref{fig:code_example} shows an excerpt of an example
Myrmics application that hierarchically processes a binary tree.
In the initialization phase (not shown in the figure), the user
creates one allocation region for the whole tree (\texttt{top},
line~\ref{src:top}) and allocates a tree, so that for each
\texttt{TreeNode *n} in a region, its left subtree \texttt{*left}
(or right subtree \texttt{*right}) is allocated in a subregion
\texttt{n->lreg} (or subregion \texttt{n->rreg} respectively).
The \texttt{root} tree node is allocated in the \texttt{top}
region. Lines~\ref{src:pragma1}--\ref{src:pragma1e} spawn one
task to process the tree, which spawns two tasks to process the
left and right subtrees recursively
(lines~\ref{src:pragma3}--\ref{src:pragma3e}). The \texttt{inout}
clause in the pragma specifies that the spawned task has both
read and write access to the \texttt{top} region.
Lines~\ref{src:pragma2}--\ref{src:pragma2e} spawn a single
\texttt{print} task to print all results. The task is dependent
on reading the whole tree, stored in region \texttt{top}. Myrmics
will therefore schedule \texttt{print} only when the
\texttt{process} task and its children tasks have finished
modifying the child regions of \texttt{top}.  When \texttt{print}
finally runs (lines~\ref{src:print}--\ref{src:printe}), it has
read-only access (as defined by the \texttt{in} pragma clause in
line~\ref{src:pragma2}) to the whole tree and can follow any
pointers freely.

Despite being a contrived example, this code highlights some
important strengths of the Myrmics programming model. The
producer-consumer task dataflow model enables software-based
cache coherency for non-coherent architectures, by transferring
task data from producer to consumer CPU cores. Scalable
runtime implementations are possible by decentralized runtime
agents who schedule these transfers. Cache-coherent,
shared-memory systems can also benefit by prefetching data to the
consumer CPU cache. Regions allow the programmer to dynamically
change parts of pointer-based structures, \textit{e.g.} to
allocate or free nodes. A task can be spawned by declaring the
region as a single dependency argument. The runtime system
guarantees that all objects (and sub-regions) in the region will
be accessible to the task code when it is executed. This not only
enhances the programming expressiveness, but also allows the
runtime system to optimize for spatial locality by keeping
objects in the same region packed close together. Moreover, using
regions to partition pointer-based data structures allows all
objects to use direct pointers. In contrast, the few programming
models and runtimes that do support pointer-based structures,
like UPC~\cite{upc,bupc}, resort to ``fat'' software pointers,
\textit{i.e.} software identifiers to metadata. Contrary to
direct pointers, such runtime systems have to mediate in order to
dereference the fat pointers, which leads to increased overhead
per such traversal.

\section{Background: Target Platform}
\label{sec:platform}

\begin{figure}
  \centering
  \includegraphics[width=0.7\columnwidth]{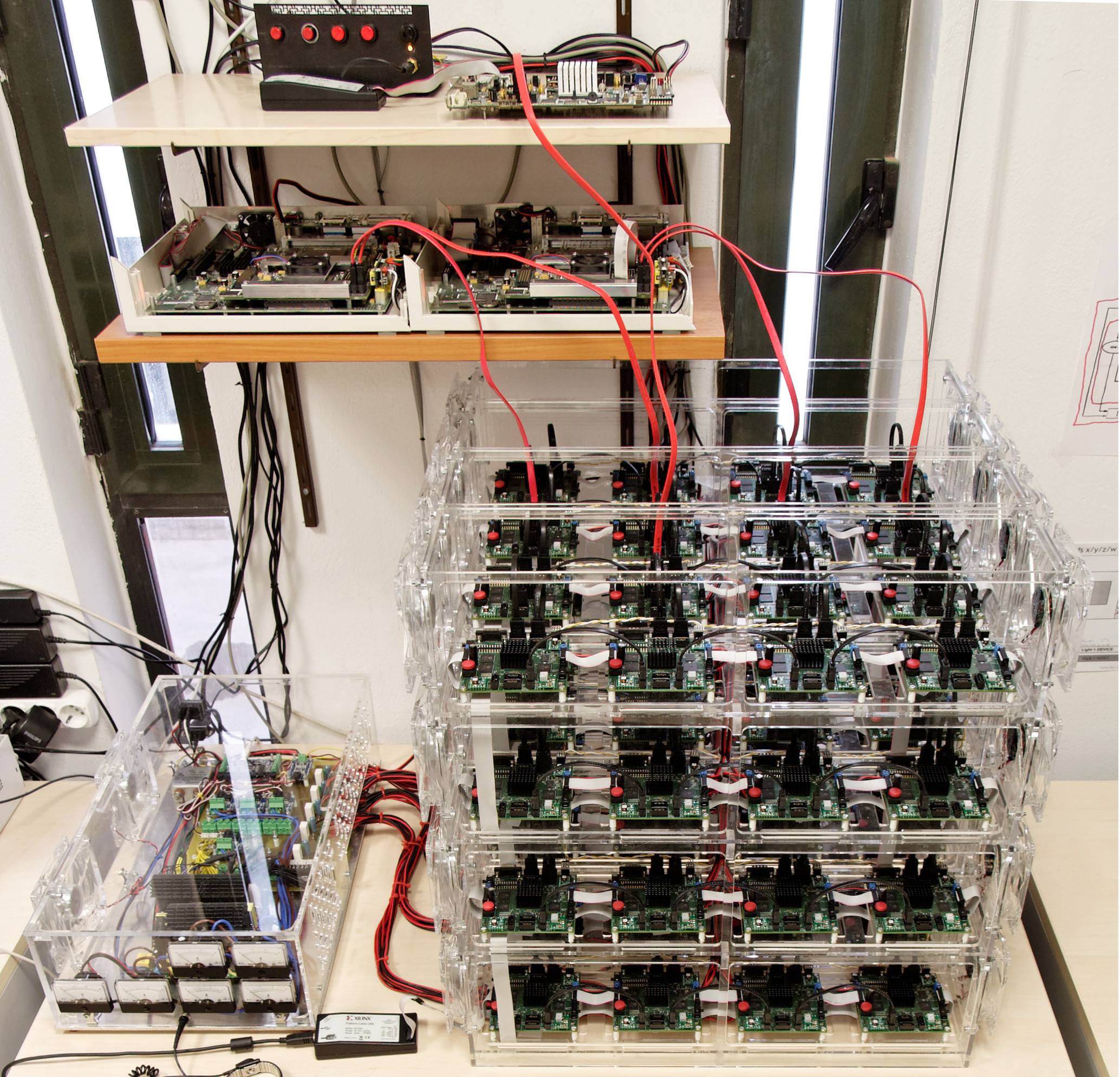}
  \caption{\footnotesize
    The 520-core heterogeneous prototype platform. Sixty-four
    octo-core Formic boards are organized in a 3D-mesh inside a
    Plexiglas cube (bottom right). Two quad-core ARM Versatile
    Express platforms (upper middle) are connected to the Formic
    cube. A modified PC power supply unit (bottom left) and a
    microcontroller-based box (top left) power and control the
    system, enabling remote access. The system models a future
    single-chip manycore processor with 512 Xilinx MicroBlaze
    (32-bit, slow, in-order) cores and 8 ARM Cortex-A9
    (32-bit, fast, out-of-order) cores. MicroBlaze cores run the
    application code (task execution) and ARM cores run
    the Myrmics runtime system, which resolves task dependencies,
    schedules tasks and keeps the caches in a coherent state.
  }
  \label{fig:prototype}
\end{figure}

One of our primary goals is to explore the scalability of
task-based runtime systems in emerging manycore CPUs. As
processors with hundreds of cores are not available today,
researchers either use architectural simulation tools like
gem5~\cite{gem5} or adapt their code to run in clusters. We find
that the simulation is very slow and therefore unsuitable to run
a runtime system and an application for hundreds of cores. Using
clusters is a far better solution; \textit{e.g.} the Legion
runtime is based on GASNet~\cite{gasnet}, a communication layer
that supports many platforms, including clusters. However,
communication latencies in clusters are orders of magnitude
greater than on-chip core-to-core communication. Because of this,
a cluster has different architectural characteristics from a
manycore chip, and consequently the conclusions derived from
clusters are not directly applicable to manycore chips. For
example, latency issues fundamentally affect the minimum size of
the application tasks. Typically, cluster-based runtime systems
employ very coarse-grained tasks to hide the communication
latency, which also helps not to stress the runtime system with a
large number of concurrent tasks. However, many fine-grain tasks
are able to harness much more parallelism; future runtime systems
should be capable of supporting them efficiently.

For all these reasons, we develop Myrmics to run on a custom
FPGA-based prototype system, pictured in
Fig.~\ref{fig:prototype}. The system contains a total of 520
CPU cores in a tightly-coupled 3D-mesh. The custom hardware is
tuned to emulate very low-latency communication, as if all 520
cores were inside a single chip. The system is heterogeneous,
featuring 8 ARM Cortex-A9 cores and 512 Xilinx MicroBlaze cores.
We use the few, fast, out-of-order Cortex cores to run the
Myrmics runtime, including the 
control-heavy scheduling and dependency-analysis, and the
many, slow, in-order MicroBlaze cores to execute the application
tasks. Runnemede~\cite{runnemede}, a recent paper from
Intel, suggests that a similar heterogeneous manycore
architecture, where a task-based runtime runs on fast cores
and tasks execute on the slow cores is optimal for energy
consumption. Production ARM-based system-on-chips already benefit
from similar hardware architectures~\cite{big_little}. These
views reinforce our intuition for a heterogeneous,
manycore platform as a viable prediction for the near future.

The hardware platform is not fully coherent. Specifically, the
ARM cores are fully cache-coherent in groups of four. Each ARM
core has private L1 (32-KB instruction and 32-KB data) caches,
and shared 512-KB L2 cache and 1-GB DRAM. The MicroBlaze cores
have private L1(4+8 KB) and L2(256 KB) caches and share an 128 MB
DRAM in groups of eight, but their L2 caches are not coherent.
Hardware DMA engines transfer data among cores, ensuring local
cache coherency between source and destination. The software
needs to keep caches coherent. This is done in principle by
transferring the up-to-date copy of produced data to consumer
tasks using DMAs, before the consumer tasks can start executing.
Also, the runtime does not depend on shared memory constructs and
locking; explicit messages are exchanged among cores for
coordination. Various hardware mechanisms are in place to assist
with message-passing, completion detection and
network-on-chip-specific behavior. Communication latencies have
been tailored to emulate single-chip manycore processors, such as
the Intel SCC~\cite{scc} prototype. A full DMA operation can be
started in 24 CPU clock cycles, a core-to-core round-trip message
time costs from 38 (nearest core) to 131 (farthest core) clock
cycles and an all-worker barrier (512 cores) is feasible in just
459 cycles. A more detailed analysis and benchmarking of the
hardware prototype can be found in\footnote{\ These references
describe the 512-core MicroBlaze system. A journal paper which
describes and benchmarks the 520-core Heterogeneous system has
been submitted in September 2012 and is under
revision.}~\cite{formic_fccm12,formic_tr}.

\section{Motivation for Key Design Choices}
\label{sec:choices}

\begin{figure}
\centering
  \subfloat[]{
    \label{fig:core_tree}
    \includegraphics[width=0.7\columnwidth]{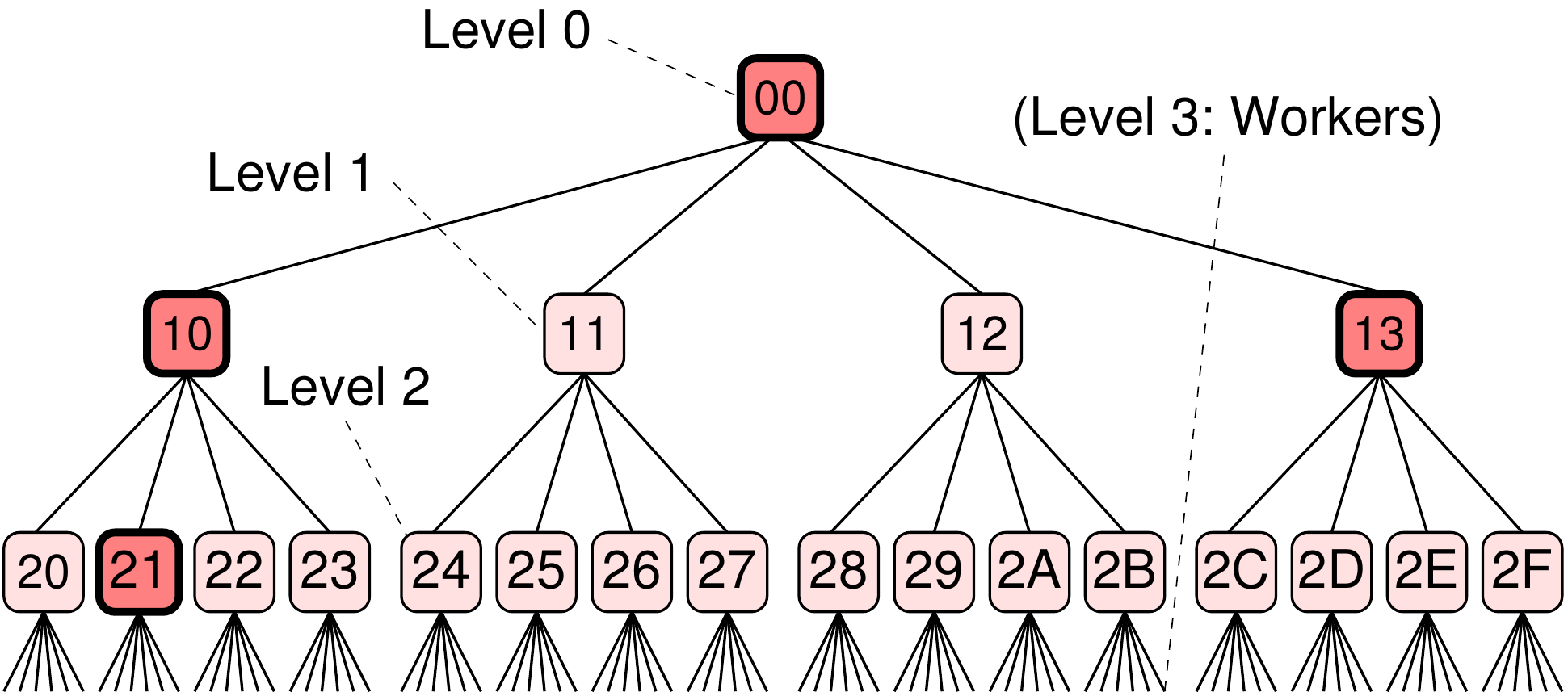}
  }
  \hfill
  \subfloat[]{
    \label{fig:region_tree}
    \includegraphics[width=0.7\columnwidth]{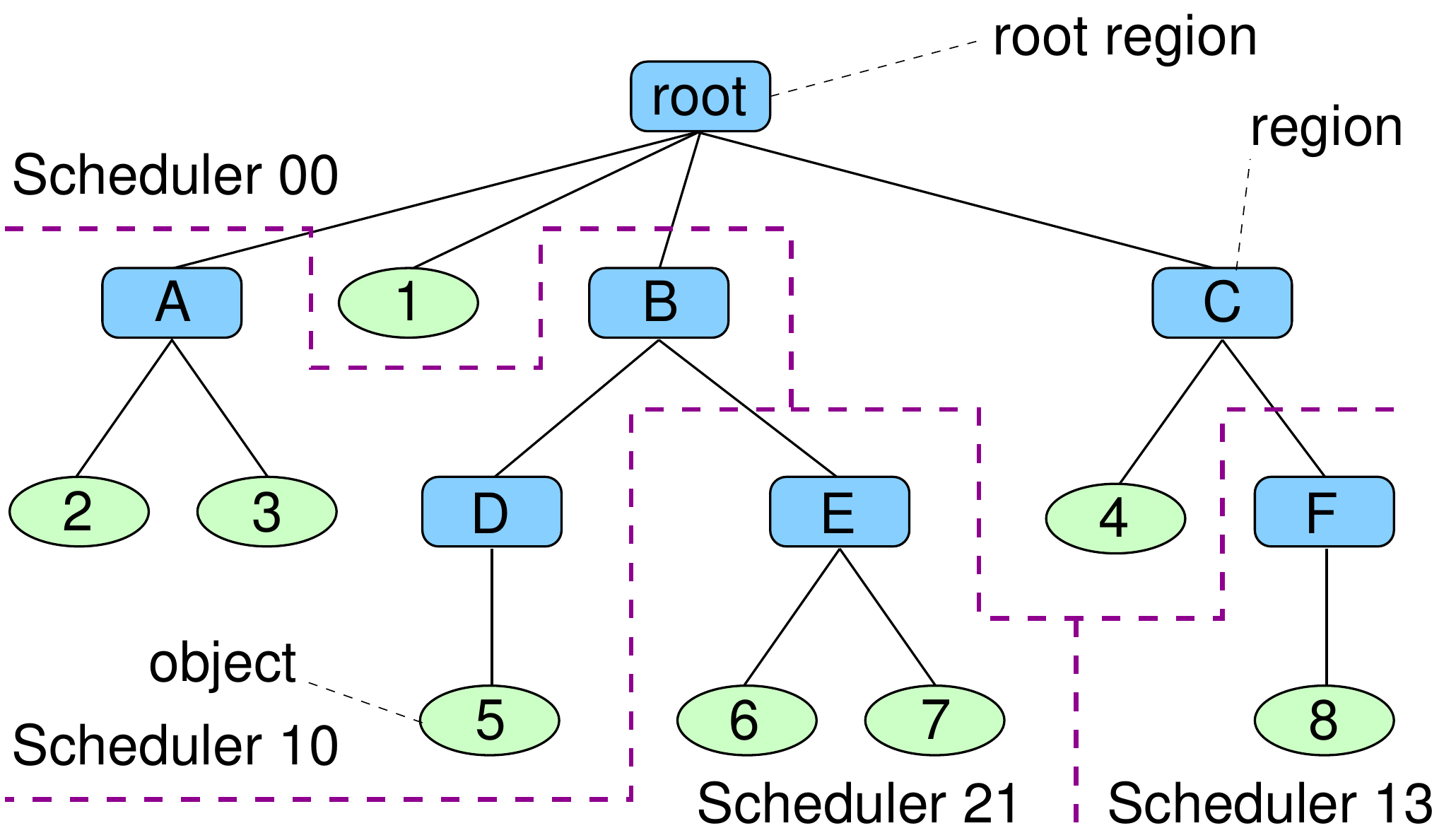}
  }
  \caption{
    Scheduler/worker tree hierarchy (a). 
    Distributing a global region tree to multiple schedulers (b).
  }
\end{figure}

\paragraph{Core specialization}
We follow the prediction made for manycore processors, that
CPUs must specialize for certain
roles~\cite{fos,last_millennium}. In Myrmics, cores become either
\textit{schedulers} or \textit{workers}. Schedulers run the main
runtime functions, like memory allocation, dependency analysis
and task scheduling. Workers just execute the tasks that
schedulers instruct them to. The per-core specialization allows
for several advantages. It improves cache efficiency, as the data
working set is smaller: a core either executes runtime code and
has in its caches runtime data structures, or executes
application code and has in its caches application data
structures.  In heterogeneous processors, assigning
control-intensive code to stronger cores and data-intensive code
to weaker cores also improves energy efficiency~\cite{big_little}
and enables many more cores to be active when operating in a
fixed power budget~\cite{runnemede}.

\paragraph{Hierarchical organization}
Schedulers and workers communicate strictly in a tree-like
hierarchy (see Fig.~\ref{fig:core_tree}). Workers form the leaves
of the tree and exchange messages only with their designated
parent schedulers. Mid-level schedulers communicate only with
parent or children schedulers. The tree root is a single
top-level scheduler. The metadata for memory objects and regions
are split among the schedulers, depending on their parent-child
relationship (also see next paragraph). We choose this setup for
three reasons. First, it allows for fast message passing. When
communication is limited to a small number of peers, we can
employ predefined per-peer buffers to safely push messages and
avoid rendezvous-like round-trips (more details are given in
section~\ref{sec:low_level}). This style of communication also
exploits any structure that may exist in the interconnection
network. Second, hierarchical structures scale well and avoid
contention. A single Myrmics scheduler keeps metadata for only a
part of the tree of memory regions and objects in the
application. Schedulers cooperate to manage memory, analyze
dependencies and schedule tasks. To serve a request, the number
of schedulers that will be involved is at most logarithmic to the
total core count. Third, hierarchy enhances data locality. A
small group of workers communicating with one or few levels of
local schedulers can solve a part of the problem isolated from
the rest of the cores. More importantly, locally spawned tasks
manage to keep all related data close to the group. This is
possible with such a hierarchical setup, but impossible with
non-hierarchically distributed data structures (such as
distributed hash tables), which would involve arbitrary cores
located anywhere on the chip.

\paragraph{Memory-centric load distribution}
A final design choice affects how the schedulers balance the load
of allocation, dependency analysis and scheduling among them. We
choose to follow a memory-centric way. Objects and regions are
assigned to the hierarchy of scheduler cores upon creation,
depending on the relationship defined by the user application,
level hints from the user as well as load-balancing criteria
(more details in section~\ref{sec:alloc}). Once assigned, they
stay on these schedulers until freed by the
application\footnote{\ We have done some preliminary analysis for
object and region migration, but we do not implement these
mechanisms in Myrmics for the moment.}. Dependency analysis for
arguments is performed by exchanging messages among the
schedulers that are responsible for the objects and regions that
comprise the memory footprints of a task. This choice has some
advantages and disadvantages. On the positive side, the user can
intuitively reason about the application decomposition by using a
hierarchy of regions. Deeper regions are mapped to lower-level
schedulers and tasks spawned to local workers, keeping data close
and reducing control message exchanges to a minimum. On the
negative side, high-level schedulers may not be utilized enough,
as the bulk of the work is performed by lower-level ones. As we
target processors with more than hundreds of cores, we consider
this to be a fair trade-off between system-wide application data
locality \textit{vs.} distribution of scheduler load.

\section{The Myrmics Runtime System Design}
\label{sec:design}

\subsection{Application Programming Interface (API)}

Fig.~\ref{fig:api} lists the Myrmics runtime system API. A
programmer may either use this interface directly to write
applications, or employ a compiler. We use a modified version of
the SCOOP compiler~\cite{scoop_pact12} to translate
pragma-annotated C source code (such as the example in
fig.~\ref{fig:code_example} to plain C code with calls to the
Myrmics API. We give an overview of the interface here. Formal
semantics and proofs for determinism and serial equivalence can
be found in~\cite{progr_model_mspc11}.

\begin{figure}
\begin{lstlisting}[numbers=none]
// Region allocation
rid_t sys_ralloc(rid_t parent, int lvl);
void sys_rfree(rid_t r);

// Object allocation
void *sys_alloc(size_t s, rid_t r);
void sys_free(void *ptr);
void sys_realloc(void *old_ptr, size_t
                  size, rid_t new_r);
void sys_balloc(size_t s, rid_t r,
                  int num, void **array);
// Task management
#define TYPE_IN_ARG            (1 << 0)     
#define TYPE_OUT_ARG           (1 << 1)       
#define TYPE_NOTRANSFER_ARG (1 << 2)     
#define TYPE_SAFE_ARG          (1 << 3)     
#define TYPE_REGION_ARG        (1 << 4)     

void sys_spawn(int idx, void **args, 
                int *types, int num_args);
void sys_wait(void **args, int *types, 
               int num_args);
  \end{lstlisting}
  \caption{The Myrmics API}
  \label{fig:api}
\end{figure}

The user allocates a new region with \texttt{sys\_ralloc()}.
The call returns a unique, non-zero {\em region ID} (of type
\texttt{rid\_t}), which represents the new region. A region is
created under an existing parent region or the default top-level
root region, represented by the special region ID $0$. A level
hint (\texttt{lvl}) informs the runtime on how deep the new
region is expected to be in the application region hierarchy, so
it can be assigned to a scheduler which is appropriately deep in
the core hierarchy. A region is freed using the
\texttt{sys\_rfree()} call, which recursively destroys the
region, all objects belonging to it and its children regions.
A new object is allocated by the \texttt{sys\_alloc()} system
call, returning a pointer to its base address. The object may
belong to any user-created region or the default top-level root
region. Objects are destroyed by the \texttt{sys\_free()} call
and can also be resized and/or relocated to other regions by the
\texttt{sys\_realloc()} call. In order to reduce worker-scheduler
communication traffic induced by memory allocation calls, we
provide the \texttt{sys\_balloc()} call, which allocates a number
of same-sized objects in bulk and returns a set of pointers. This
call minimizes communication for common cases like the allocation
of table rows.

A running task spawns a new task by calling \texttt{sys\_spawn()}
and specifying an index to a table of function pointers. Two
tables are passed to this call, one containing the actual task
arguments and another describing the dependency modes for them.
Each argument can have read (\texttt{IN}) and/or write
(\texttt{OUT}) permissions. For regions, the region ID is passed
as an argument and the \texttt{REGION} bit indicates it is a
region. Myrmics will not perform dependency analysis for
arguments that are marked as \texttt{SAFE}. This is useful for
passing by-value arguments (\textit{e.g.} scalar values) to
tasks, for objects that already belong to regions passed to the
task, or for cases where compiler static analysis can prove that
an argument is indeed safe to use because of other overlapping
dependencies. The \texttt{NOTRANSFER} bit indicates that although
normal dependency analysis semantics apply, the actual data will
not be used by the task, so no DMA transfer is needed. This is an
optimization that can be used in non-coherent machines to avoid
DMAs for tasks whose purpose is to spawn smaller tasks, but not
actually use any objects in a region. Finally, \texttt{sys\_wait()}
can be used inside a task that has delegated (parts of) its
regions or objects to children tasks and needs to operate again
on them. The arguments and dependency modes arrays are similar to
the ones used by \texttt{sys\_spawn()}. The call suspends the
task and resumes it when all arguments are again available
locally with the requested permissions.

\subsection{Low-Level Layers}
\label{sec:low_level}

Myrmics runs directly on the heterogeneous prototype platform
without any underlying operating system or hypervisor. The reason
for the bare-metal choice is that porting Linux to the prototype
platform is very hard, as it assumes underlying cache coherency.
The lowest Myrmics layer is the \textit{architecture-specific}
one, split into ARM and MicroBlaze parts. The cores boot and
initialize their peripherals. Small device drivers present a
unified, architecture-independent interface to the higher layers
for operations such as cache management, communication and
synchronization primitives, interrupts, timers and serial port
I/O.
A \textit{kernel toolset} layer provides a user library and a set
of commonly needed utilities for programming the rest of Myrmics.
It consists of a number of functions for common data structures
(lists, tries, hash tables), a small string library, printing
functionality and a basic math library.

A \textit{Network-on-Chip (NoC)} layer implements
fast communication among scheduler and worker cores. The NoC
layer achieves effective core-to-core communication which is more
suitable for many-core processors than the heavyweight
communication libraries used in clusters. It provides two
primitives, \textit{messages} and \textit{DMA transfers}.
Messages are used to transfer control information among
schedulers and workers. Cores exchange messages only with their
parent and children cores, as defined in the core hierarchy which
is set up during the NoC layer initialization. The message size
is fixed, but programmable. We currently use a message size of 64
bytes, which coincides with a single hardware cache line. We
assign a number of per-peer software buffers, where a peer can
push messages using one-way hardware DMA primitives. Hardware
primitives are used to implement efficient polling for incoming
messages on the receiver side and a credit-flow system for the
software buffers, so no overflow can occur under system load.
Messages are very efficient and can be processed back-to-back in
the order of 450-750 clock cycles, depending on core distance and
buffer availability. The NoC layer provides software-supervised
DMA transfers and it can accept multiple DMA transfers in groups.
Since a DMA transfer can fail if a queue is full at the
destination core, the NoC layer restarts the failed DMA transfers
and when all transfers in a group are successfully completed it
notifies the upper software layer.


\subsection{Memory Management}
\label{sec:alloc}

Myrmics implements a global address space with
software-maintained coherency. We have developed and benchmarked
the Myrmics memory management subsystem in previous
work~\cite{myrmics_ismm12}, where we explain in detail how we
implement the global address space with common pointers.
Coherency issues are explained further in
section~\ref{sec:scheduling}. This section overviews the
memory management subsystem functionality. 

%

The global address space is implemented by multiple, cooperating
scheduler instances. Schedulers are connected in a tree
hierarchy, as Fig.~\ref{fig:core_tree} shows. The tree has one
top-level scheduler with a configurable number of next-level
children, depending on the number of processor cores and their
capabilities. The scheduler tree descends for some levels. Each
lowest-level scheduler core is responsible for a number of worker
cores. Cores exchange messages only with cores that are one level
above or below them in the hierarchy. Each scheduler is organized
as an event-based server. The hardware architecture supports
event-driven execution and the NoC software layer extends this
functionality by enabling a core to sleep until a new message
arrives. Scheduler cores are in a continuous loop, waiting for
new messages. We globally construct a \textit{region tree}, such
as the one shown in fig.~\ref{fig:region_tree}, based on the
relationship of user-allocated regions and objects. Each
scheduler core handles a part of the global region tree. Its
portion includes whole regions and any objects that belong to
them, but not necessarily all of their descendant regions. The
highest level of the Myrmics memory subsystem responds to
memory-related messages. If an incoming message refers to a
region handled locally, the server immediately processes it and
responds. Otherwise, the server forwards the message to its
parent or child schedulers. Reply messages from other schedulers
are intercepted if they refer to pending actions for which the
local scheduler awaits reply, otherwise they are forwarded to the
original requesters. We also support reentrant events with saved
local state for more complex situations in which we can handle
part of the request locally or the final response should be
assembled from multiple remote responses. We divide work between
schedulers based on the regions that are local to them. Worker
access to objects and regions that are not local to the
lowest-level scheduler incurs inter-scheduler communication, so
that the scheduler that is responsible for the region can handle
the request. The Myrmics scheduling subsystem primarily attempts
to minimize this cost. We assign regions to schedulers using both
a level hint from the programmer and load-balancing criteria. We
use the hint to estimate the ``vertical'' positioning of a region
on the scheduler hierarchy and load balancing to determine
the ``horizontal'' positioning; a non-leaf scheduler that must
assign a new region to one of its children does so by selecting
the one with the lowest region load. Fig.~\ref{fig:region_tree}
shows an example of how the region tree is split among the
schedulers in the hierarchy of Fig.~\ref{fig:core_tree}.
Schedulers use tries to track which region IDs and address
ranges belong to which children schedulers. They also
periodically exchange upstream load information messages,
whenever the previous reported load differs by a configurable
threshold.

The Myrmics memory allocator uses an underlying slab allocator to
support hierarchical regions that are local to a scheduler
instance.  We use a new slab pool to build each local region when
it is created. Packing region objects in dedicated slabs helps to
isolate them from other regions and to enable communication on
slab-based quantities. Allocating a new slab pool per region
increases fragmentation, because partially filled and
preallocated empty slabs are dedicated for the new region. We
consider this trade-off to be acceptable since many future object
allocations in the region will happen quickly and will be
compacted with other region objects, increasing communication
efficiency and locality of region objects. We use an adaptive
mechanism that is based on watermarks to control the limit of
external fragmentation. Local regions trade slabs based on their
utilization, before the scheduler requests new pages from its
parent. This policy reduces communication and balances increased
locality of region objects with increased fragmentation.
The underlying slab allocator manages the dynamic allocation and
freeing of memory objects of any size organized in packed groups
of same-sized objects. We tune the slab allocator to the size of
the 64-B cache lines in our target platform. We design the slabs
so that their metadata are carefully separated from the data,
which increases the efficiency of hardware transfers and
facilitates moving whole regions with few operations. The
allocator uses a slab size of 4~KB as the basic unit inside a
scheduler a chunk of memory and a 1-MB page size as the
basic unit which schedulers trade free address ranges to
implement a global address space.

\subsection{Dependency Analysis}
\label{sec:dep}

\begin{figure}
\centering
  \subfloat[]{
    \label{fig:dep_queues}
    \includegraphics[width=0.7\columnwidth]{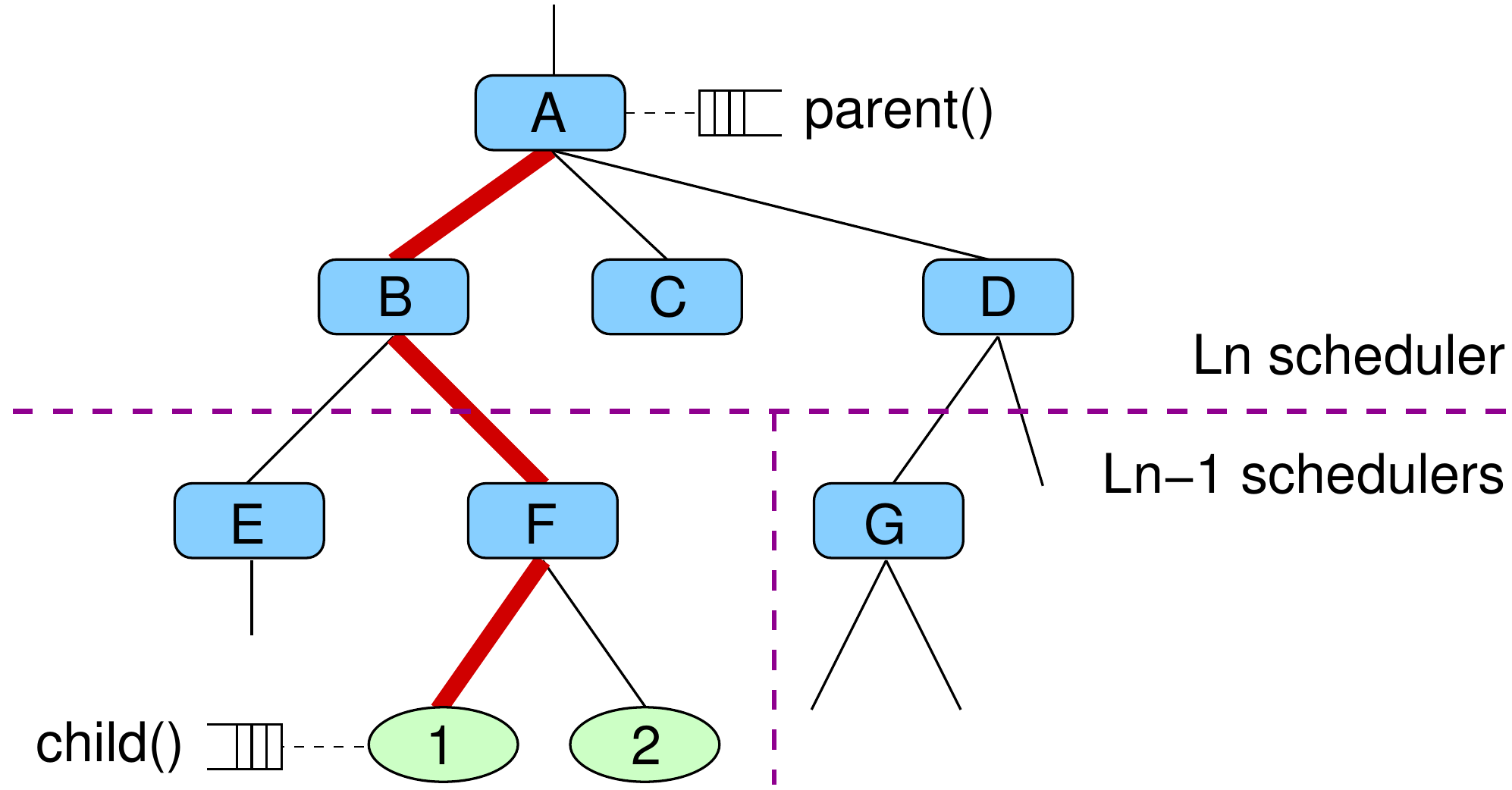}
  }
  \hfill
  \subfloat[]{
    \label{fig:dep_cnt}
    \includegraphics[width=0.7\columnwidth]{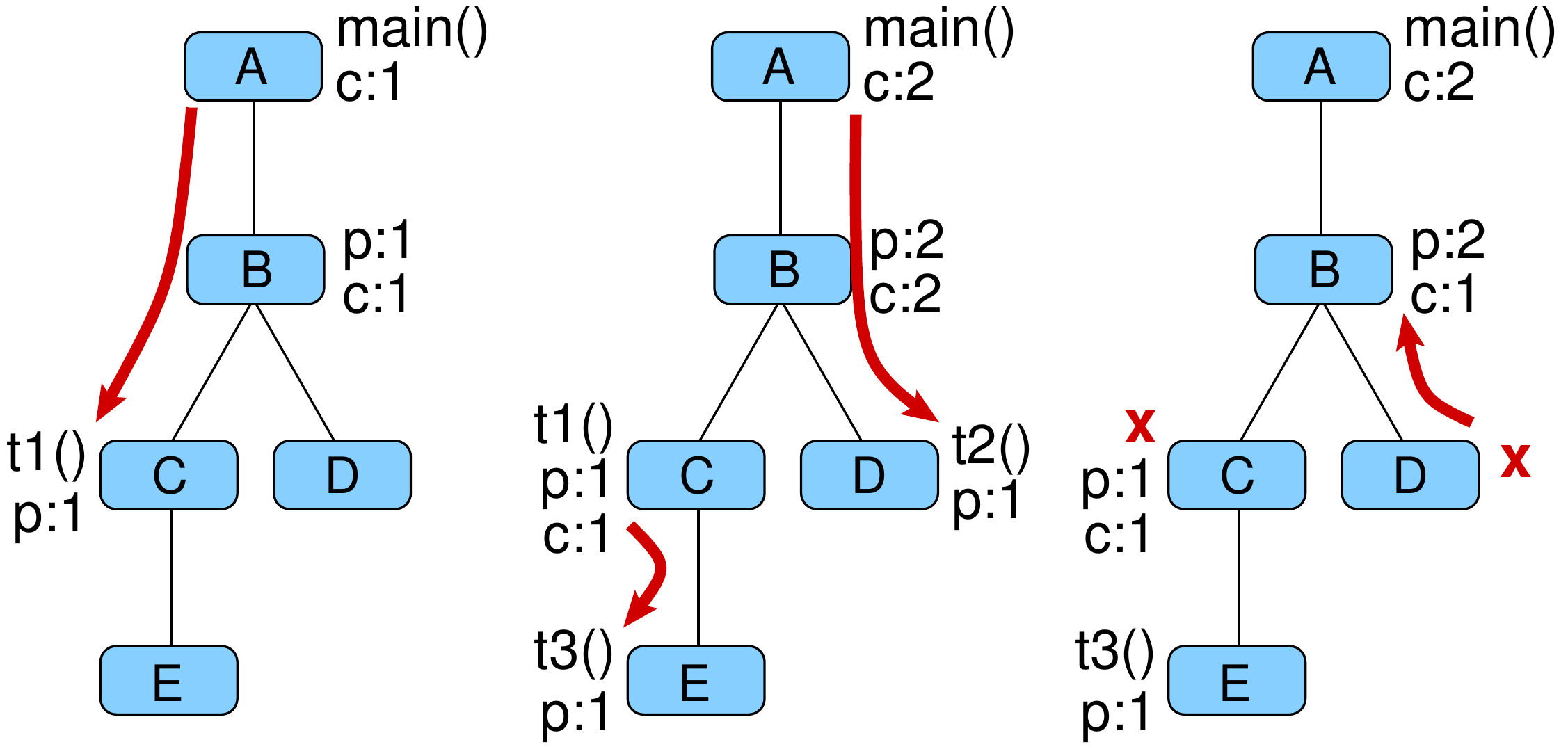}
  }
  \caption{
    Dependency queues and region traversal path (a).
    Parent/children dependency counters (b).
  }
\end{figure}

The dependency analysis subsystem of Myrmics is based upon the
main memory abstractions, objects and regions. We augment their
metadata to include \textit{dependency queues}, which are
in-order lists of tasks waiting for access. A task is
dependency-free and ready to be scheduled when it is at the head
of the dependency queues for all its arguments; in the case of
regions, no children regions should be busy as well, as
explained below. Fig.~\ref{fig:dep_queues} shows a
part of a region tree split among three schedulers. To illustrate
the dependency analysis process, assume that a parent task,
\texttt{parent()}, is already dependency-free, scheduled and
running, having a single argument which is region A. Task
\texttt{parent()} is at the head of A's dependency queue.  Assume also
that it spawns a child task, \texttt{child()}, which has a single
argument, object 1. To successfully maintain
the programming model semantics~\cite{progr_model_mspc11}, the
dependency analysis subsystem must traverse the region tree from
the point \texttt{parent()} is enqueued towards the child
argument and enqueue \texttt{child()} there. This is the red
(thick) path A$\rightarrow$B$\rightarrow$F$\rightarrow$1 in the
figure. If any dependency queue is non-empty (or if any region
children are busy) during the traversal, the process stops and
\texttt{child()} is enqueued at the end of the local queue
instead, indicating its final target is object 1 and not the
local region. \textit{E.g.}, if another task \texttt{child2()}
was at the head of F's queue, it would imply that
\texttt{child2()} should run on the whole region F before
\texttt{child()} is allowed to run using object 1, which is a
part of F. In this case, the traversal will resume when all
previous tasks in the queue are finished.

To know if a region has any children regions or objects with any
tasks in their queues we keep several software counters per
region. Whenever an argument traversal as the one we described
passes through a region, we increment a counter in the region to
indicate that one of its children has a pending task.
Fig.~\ref{fig:dep_cnt} shows a simplified example. At the
left-hand part, \texttt{main()} who owns region A spawns
\texttt{t1()} to work on region C. Counter ``c'' (for ``child'')
is incremented in regions A and B to note there's one child
enqueued for a part of these regions. At the middle part of the
figure, this happens again when \texttt{main()} spawns
\texttt{t2()} to work on region D, and \texttt{t1()} spawns
\texttt{t3()} to work on region E. Note that now the child
counter in region B has two children pending. When a task
finishes, the next task waiting in the dependency queue of
each task argument is marked as ready. If the queue is empty, no
more tasks are waiting for this argument and the parent region is
notified that one of its children has finished. The parent region
decrements its child counter. When the counter reaches 0, it
means that all its children have finished and that the next task
waiting for the whole region can now proceed. In the right-hand
part of Fig.~\ref{fig:dep_cnt}, \texttt{t1()} and \texttt{t2()}
both finish and their queues are empty. Region C child counter is
non-zero, as it has one more child operating on a part of it
(\texttt{t3()} on region E), so nothing happens. Region D child
counter is zero, and so parent region B is notified and
decrements its child counter, which is now 1. Nothing more happens
as B still waits for one more child (region C, which is at this
time delegated to \texttt{t3()} on region E). Myrmics uses separate
child counters to indicate read/write or read-only dependencies, so we
can optimize for multiple tasks to have access to read-only arguments.

As fig.~\ref{fig:dep_queues} shows, the region tree path between
a parent and a child task may be split between two (or even more)
levels of schedulers. Whenever this happens, for task spawns or
task completions, we exchange a message between the boundary
schedulers with enough information to continue the operation as
needed. Task spawning is the most expensive operation, as it
requires multiple traversals. This happens because the parent
task can spawn a child which is arbitrarily deep in the region
hierarchy. In the example, \texttt{parent()} owns region A and
spawns \texttt{child()} to operate on object 1. Myrmics
schedulers keep sufficient information to locate object 1 in
$O(1)$ time, if it is in the same scheduler as \texttt{parent()},
or indicate which scheduler must be contacted next to go towards
the object. However, there is no information about which exact
regions lie in the path from A$\rightarrow$1, \textit{i.e.} B and
F in this case\footnote{\ We specifically choose not to keep such
information. because doing that would lead to a non-scalable
setup. Each time a new region or object was created, we would
have to update all regions up to the root of the region tree to
include the path towards the newborn.}. To discover the path, we
locate the target (possibly through messaging the scheduler where
it resides) and follow parent pointers until we encounter the
parent task, keeping track what are the intermediate regions we
pass through. We then begin the downwards traversal, as
described previously. 

Throughout the dependency analysis, each Myrmics scheduler
minimizes the number of messages among schedulers by considering
all task arguments simultaneously. Schedulers group necessary
communication and pack information for multiple task arguments
into as few messages as possible. We further analyze these
message exchanges between boundary schedulers to avoid race
conditions. Specifically, a hazard exists when a child boundary
region finishes its last task and sends an upward message to
notify its parent region that it is finished, while at the same
time a new task passes through the boundary parent region and
sends a message to the child to enqueue it there. We avoid this
race by employing ``parent'' counters in every region that keep
track how many enqueue requests have been received from its
parent. When the dependency queue becomes empty, the child
scheduler adds to the message towards its parent scheduler the
number of completed enqueues from this counter. The parent
compares this number to its child counter and disregards the
request to proceed to the next task if the numbers do not match.
Fig.~\ref{fig:dep_cnt} shows these counters as ``p'' (for
``parent''). As \texttt{t2()} completes (right-hand part of the
figure), the parent counter in region D has the value 1. Thus, 1
is decremented from region B child counter.

\subsection{Task Scheduling}
\label{sec:scheduling}

\begin{figure}
  \centering
  \subfloat[]{
    \includegraphics[width=0.7\columnwidth]{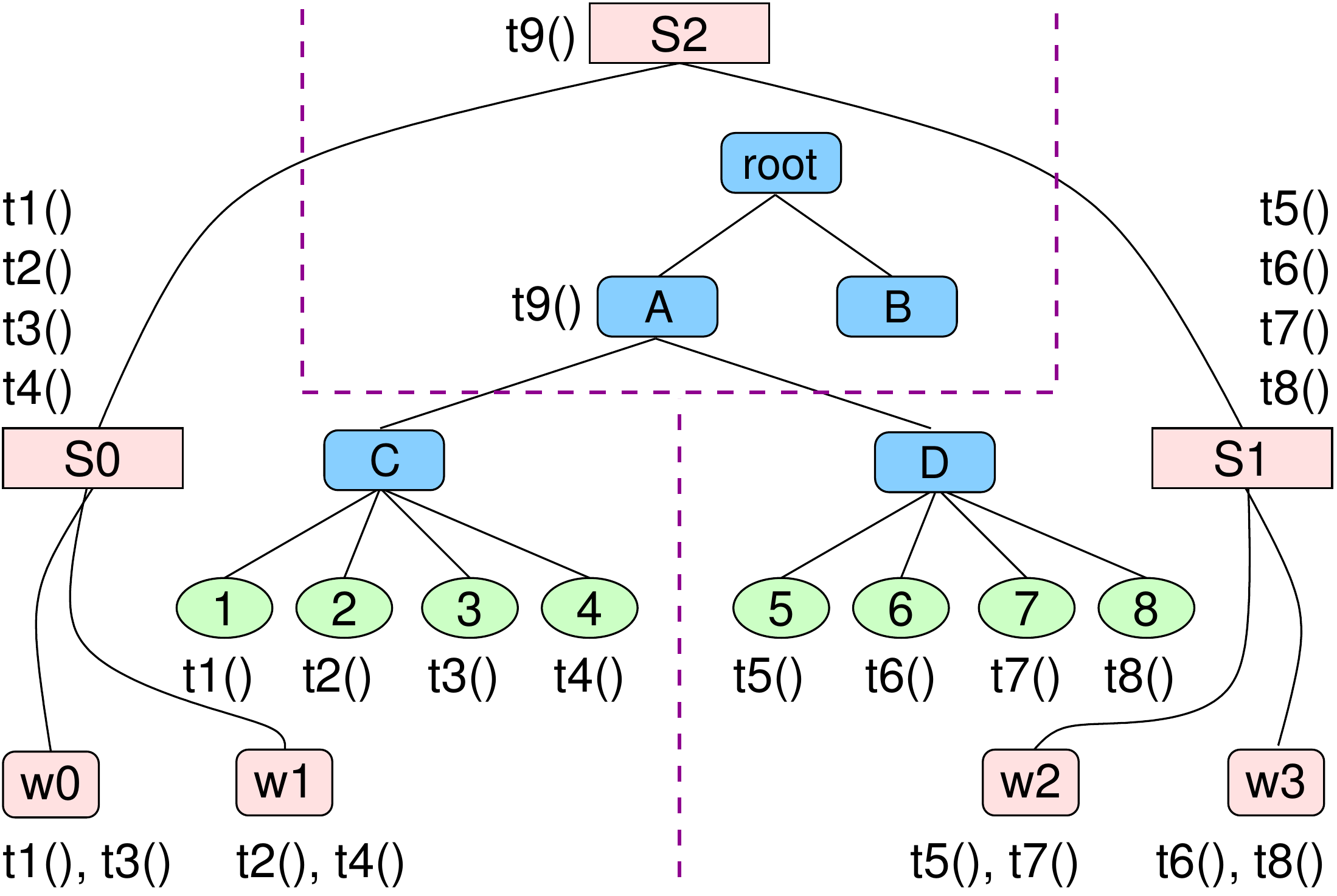}
    \label{fig:scheduling1}
  }
  \hfill
  \subfloat[]{
    \includegraphics[width=0.7\columnwidth]{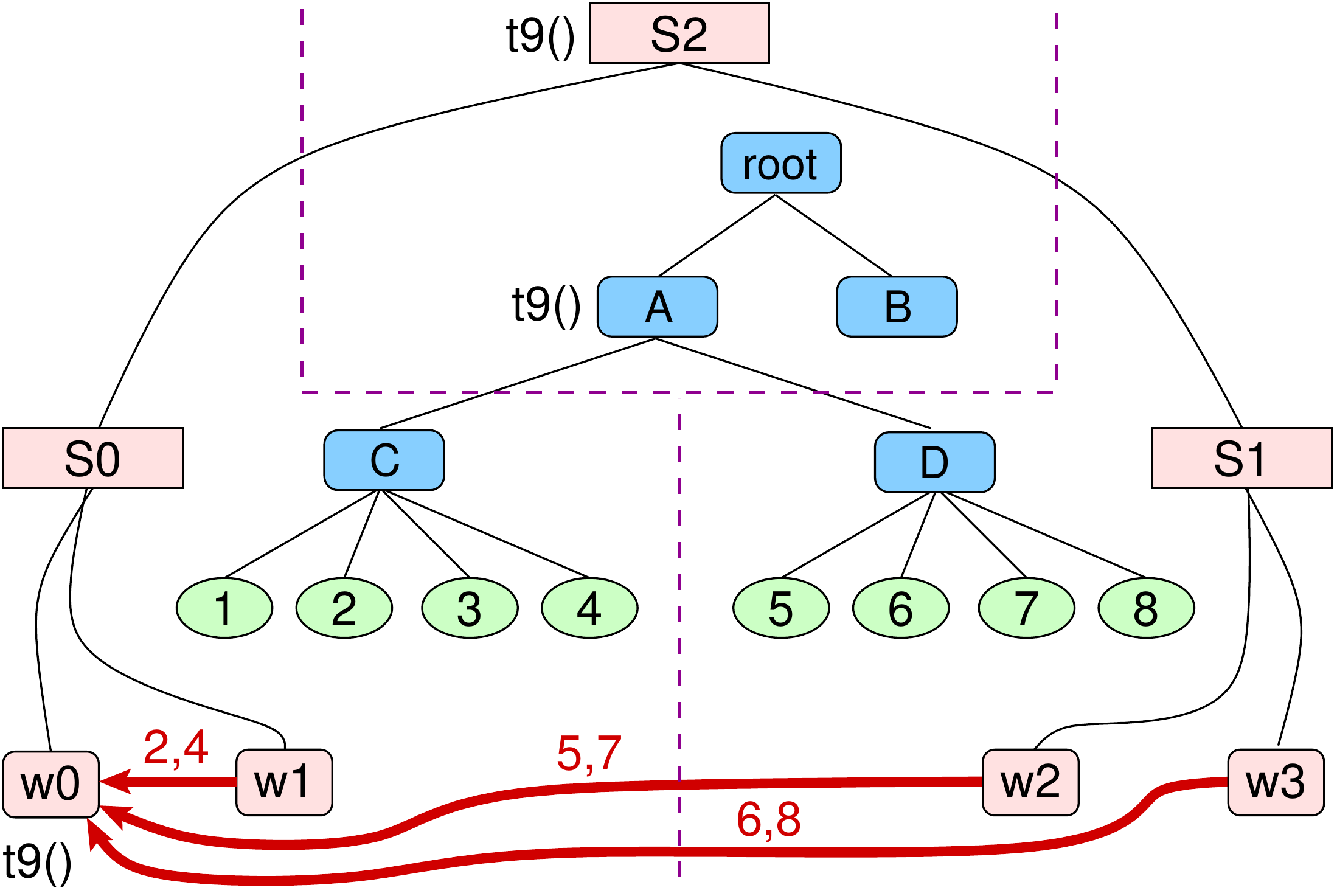}
    \label{fig:scheduling2}
  }
  \caption{
    Scheduling example. All workers working on objects (a),
    worker 0 doing a reduction (b).
  }
  \label{fig:scheduling}
\end{figure}

Each task in Myrmics is assigned to one of the schedulers, which
is responsible to monitor it until it retires. When a parent task
spawns a new child, the responsible scheduler of the parent task
handles the spawn request. The scheduler inspects the arguments
that the new task requires and has two options: either to create
the new task locally, or to delegate it to one of its children
schedulers, if it has any. We decide to delegate a new task to a
child scheduler only when all its arguments are handled by this
single child scheduler or its children. To illustrate this
concept, fig.~\ref{fig:scheduling1} shows both a region tree and
how we split it among three schedulers. Task \texttt{t1()}
operates only on object 1. Let us assume that the scheduler
responsible for \texttt{t1()}'s parent task (not shown in the
figure) is S2. Upon the creation of \texttt{t1()}, S2 observes
that all its arguments (object 1) are assigned to S0. Thus, the
creation of \texttt{t1()} is delegated to S0. Using this
memory-centric criterion to balance the task scheduling load
among schedulers is consistent with our key design choices, as
explained in section~\ref{sec:choices}. 

After a task is created, the dependency analysis subsystem takes
over to guarantee that all task objects and/or regions are safe
to be used according to requested read/write privileges. When the
task is dependency-free, it becomes ready to be scheduled for
execution. To make an informed decision, the scheduler
responsible for the task initiates a \textit{packing} operation
for all task arguments. Packing creates an optimized, coalesced
list of address ranges and sizes of all the regions/objects,
grouped by the \textit{last producer} of each such range.
The last producer of data in Myrmics is defined as the last
worker core which had write access to a specific address range.
Packing is a process carried out by the memory subsystem, which
may be hierarchical and require communication among other
schedulers lower in the hierarchy. In
fig.~\ref{fig:scheduling1}, packing region A is required to
schedule \texttt{t9()} when it becomes dependency-free. S2 will
exchange messages with S0 and S1 to pack regions C and D
respectively.

When packing of all task arguments is complete, the scheduling
begins. The total sizes per last producer are used as weights to
create a locality score, $L$: scheduling the new task to a core
which has produced a large part of the data it needs, yields a
higher locality score. We also compile a load-balancing score,
$B$, based on periodic load report messages that flow upstream in
the core hierarchy. Both $L$ and $B$ are normalized between 0 and
1024.  We combine them to create a total score, $T = pL + (100 -
p)B$, where $p$ is a policy bias percentage which we can use to
favor one of the two scores over the other. The configurable
policy bias allows exploring the producer-consumer data locality
and limit message traffic. We evaluate its effect in
section~\ref{sec:policy_eval}.  The scheduling decision may again
be hierarchical in nature. If the scheduler has children
schedulers, its decision refers to scheduling the task to the
part of the core hierarchy managed by one of its scheduling
children. The process repeats until a leaf scheduler decides
which of its workers will run the task.  The scheduler
responsible for the task dispatches it for execution towards the
chosen worker core. At the same time, if any of the task
arguments were requested for write access, it informs the memory
subsystem that from now on the last producer is the chosen
worker. 

Worker cores run a very small portion of the Myrmics runtime
system. They await messages from their parent scheduler which
dispatch tasks to be executed. Workers implement ready-task
queues to keep these task descriptors. Some task arguments may be
local to the worker core ---if it was the last producer for
them--- and others may be remote. The locality score $L$, as
discussed above, favors scheduling tasks to the same worker cores
where their arguments were last produced.  The worker orders a
group of DMA transfers for all remaining remote arguments to be
fetched from their last producers. The first task in the
ready-task queue is allowed to begin execution when the DMA group
has successfully completed.  Fig.~\ref{fig:scheduling} shows an
example. In the left sub-figure, eight tasks
\texttt{t1()}--\texttt{t8()} operate on eight different objects.
Each worker w0--w3 has two tasks in its ready queue. In the right
sub-figure, after all eight tasks have finished, task
\texttt{t9()} which will perform a reduction on the whole region
A is now dependency-free and scheduled to run on worker w0. To do
so, w0 performs DMA transfers for objects 2, 4, 5, 6, 7, 8 from
their last producers. Whenever two or more task descriptors exist
in the queue, the worker optimizes the DMA transfers by ordering
the DMA group for the second task to the NoC layer before
starting to execute the first task. This technique allows for
efficient double-buffering, as communication for the next task is
hidden by the hardware during computation of the current task.
Workers do not interrupt running tasks. If a task calls the
runtime for any reason (\textit{e.g.} to spawn a new task or do a
memory operation), the NoC layer checks for new messages and
progress with outstanding DMA transfers.

\section{Evaluation}
\label{sec:evaluation}
We evaluate Myrmics in several ways. We use a set of benchmark
applications and microbenchmark experiments to measure the overheads,
performance, scaling, locality, load-balancing, and the effect of
scheduler hierarchy configuration.

\subsection{Intrinsic Overhead}
\label{sec:overhead}

\begin{figure}
\centering
  \subfloat[]{
    \includegraphics[width=0.7\columnwidth]{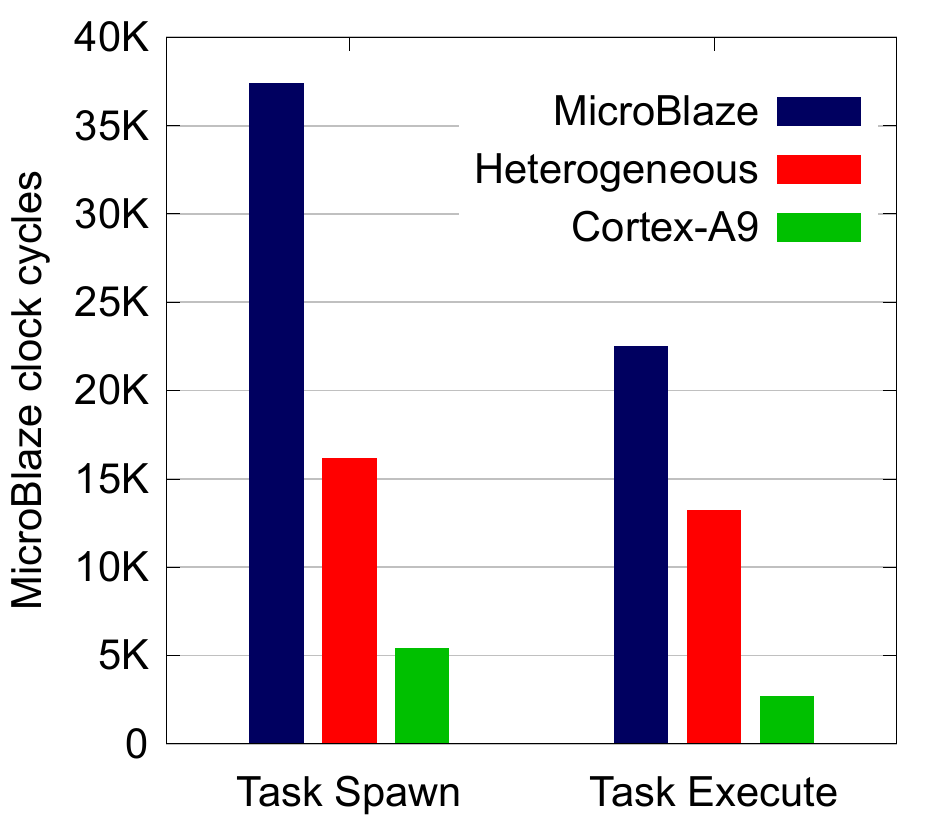}
    \label{fig:task_times}
  }
  \hfill
  \subfloat[]{
    \includegraphics[width=0.7\columnwidth]{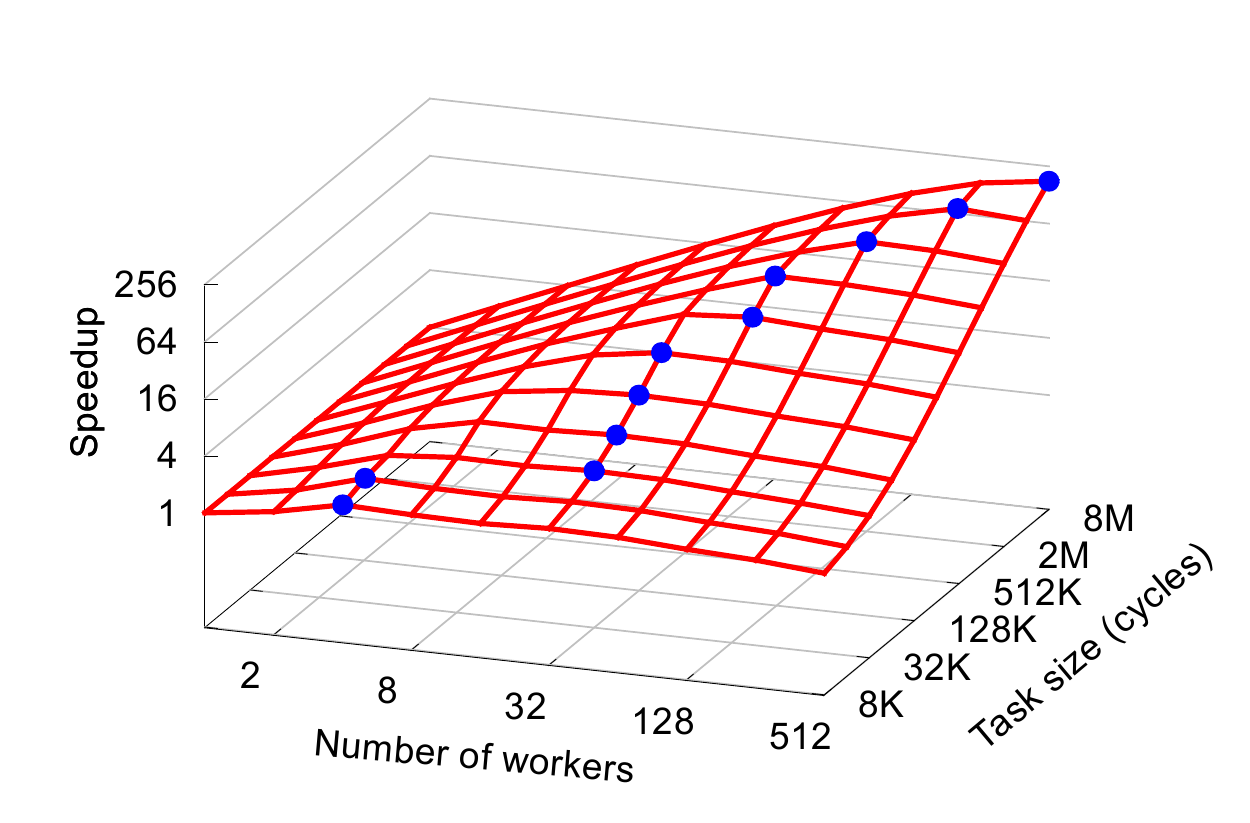}
    \label{fig:task_synthetic}
  }
  \caption{
    Time to spawn and execute an empty task (a).
    Impact of task granularity on achievable speedup (b).
  }
\end{figure}

The inherent overheads of the runtime system place a lower limit
to the task sizes that Myrmics can handle as well as how well can
the system scale.  We create a synthetic microbenchmark that
spawns 1,000 empty tasks with the same one object as an argument.
We use a single scheduler core and a single worker core.  We
measure the time required for the worker to spawn all the
tasks.  We also measure the time required for the worker to
execute all spawned tasks in order, as there is no other worker
in the system.
Fig.~\ref{fig:task_times} shows the results, normalized to the
time for a single task (we divide the total times by 1,000). We
do this experiment in three modes: with the scheduler and worker
being MicroBlaze cores (left/dark blue bars), with a Cortex-A9
scheduler and a MicroBlaze worker (middle/red bars) and with both
cores being Cortex-A9 (right/green bars). To have a common time
reference, all results are measured in MicroBlaze clock cycles.
We observe that the two CPU flavors have approximately a
7-8$\times$ difference on running time. Note that we use the
heterogeneous setup for all the evaluation that follows (except
for section~\ref{sec:deeper}), as it is the most complex and
interesting of the three. This microbenchmark shows that to spawn
an empty task with one argument, a Myrmics application needs 16.2
K cycles and to execute such a task it needs 13.3 K cycles. These
times represent the minimum overhead to execute all the
appropriate runtime functions on the worker and scheduler cores,
as well as all their communication.

We create another microbenchmark to reproduce and measure the
single-master bottleneck in Myrmics. We use one scheduler core
and a variable number of worker cores, from 1 to 512. We let the
main task spawn 512 independent tasks, each one operating on a
different object. The children tasks wait for a programmable
delay before they return.  Fig.~\ref{fig:task_synthetic} shows
the results. Axes X and Y represent our configuration (number of
workers and the task size) and axis Z shows the achieved speedup
\textit{vs.} the single-worker configuration.  We observe that
the achievable speedup for a given number of workers is limited
by the task size. Bigger tasks need scheduler interaction less
frequently, which makes the single scheduler more available to
other workers.  Notice that there is an optimal number of workers
for a given task size (filled circles in the figure).
Near the optimal point the scheduler processes tasks and fills
the worker queues fast enough so that the workers are always
busy. Adding more workers degrades performance, because there are
more events for the scheduler to process (task completions) in
less time, while new tasks always go to new, empty workers. We
approximate the optimal number of workers as the division of the
task size by the intrinsic overhead per task (16.2 K cycles). The
experiment verifies this: \textit{e.g.} for 1 M task size
Fig.~\ref{fig:task_synthetic} shows the optimal point to be 64
workers, near the computed 64.7 (1 M / 16.2 K). Finally, note
that for a given number of workers, bigger tasks always lead to
better speedup ---and asymptotically towards the perfect speedup.
These observations are also valid for hierarchical task spawning,
as they depend on the inherent Myrmics overheads.

\subsection{Scaling}
\label{sec:scaling}

We evaluate Myrmics scalability using six benchmarks, five
computation kernels and one application.
For each benchmark described below, we compare
a baseline MPI implementation to two Myrmics variants: one with a
single scheduler and one with a multiple schedulers in a
two-level hierarchical configuration.  Hierarchical Myrmics
benchmarks first use regions to decompose the computation to
coarse tasks, each of which then spawns finer tasks with object
arguments.  In all MPI and Myrmics setups we hand-select the
assignment of MPI ranks and Myrmics workers/schedulers, so that
they map as well as possible to the physical topology of the 3D
hardware platform.  To demonstrate fine-grain task parallelism,
we use task sizes down to 1~M clock cycles, which would translate
to 0.25~ms tasks in a server with 4.0~GHz CPUs. To measure strong
scaling, we use a fixed problem size and split it into
variable-sized tasks, according to the available workers. We
decompose the problem to a few (2--3) tasks per worker and per
computation step and use datasets and task sizes that fulfill
these constraints.  To measure weak scaling, we use minimum-sized
tasks and grow the problem size according to the available
workers. The algorithms for the MPI and Myrmics variants are
the same. We use non-trivial, optimized implementations
that double-buffer the data structures, overlap computation with
communication steps and perform broadcasting/reductions using
scalable (\textit{e.g.} tree-like) mechanisms. For each data
point, a Myrmics worker and an MPI core perform the same amount
of computation.

\begin{figure*}
\centering
  \subfloat[Jacobi Iteration (strong)]{
    \includegraphics[width=0.315\textwidth]{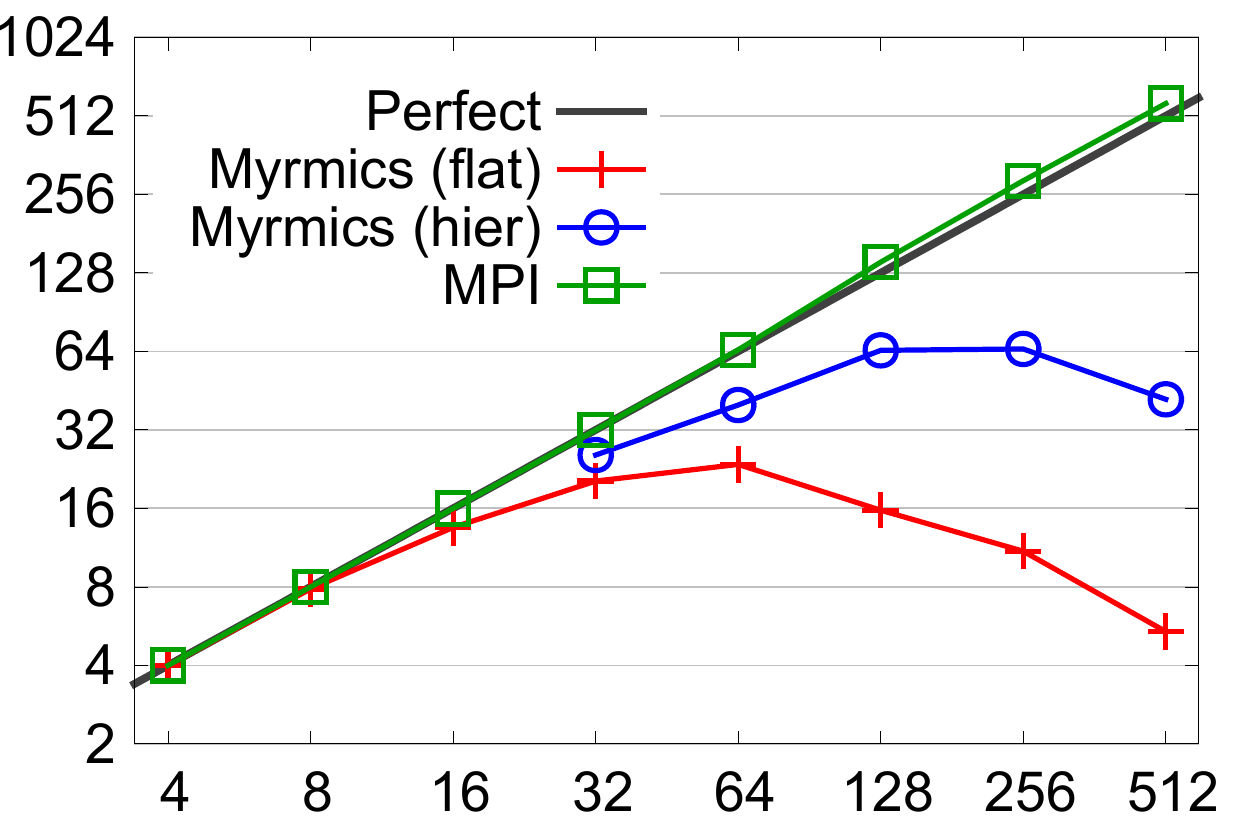}
    \label{fig:jacobi_strong}
  }
  \hfill
  \subfloat[Raytracing (strong)]{
    \includegraphics[width=0.315\textwidth]{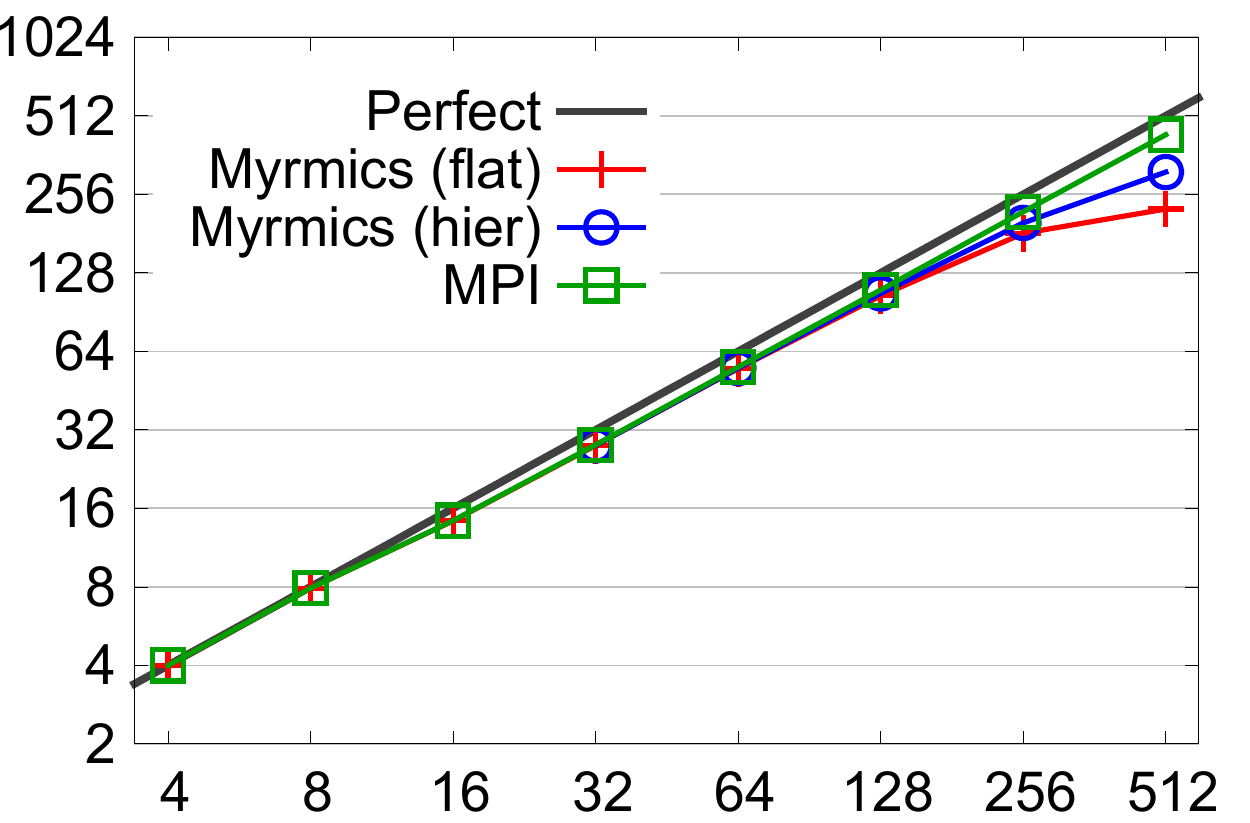}
    \label{fig:cray_strong}
  }
  \hfill
  \subfloat[Bitonic Sort (strong)]{
    \includegraphics[width=0.315\textwidth]{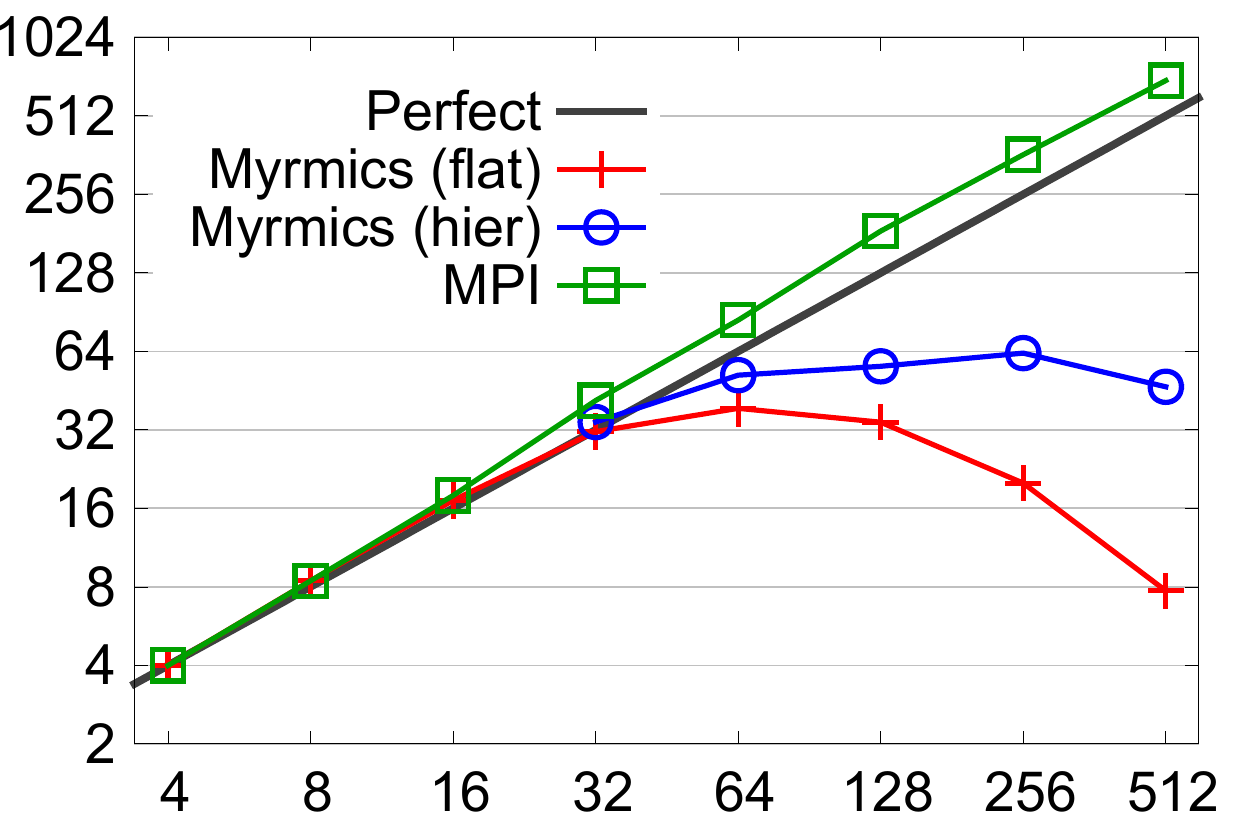}
    \label{fig:bitonic_strong}
  }

  \subfloat[K-Means Clustering (strong)]{
    \includegraphics[width=0.315\textwidth]{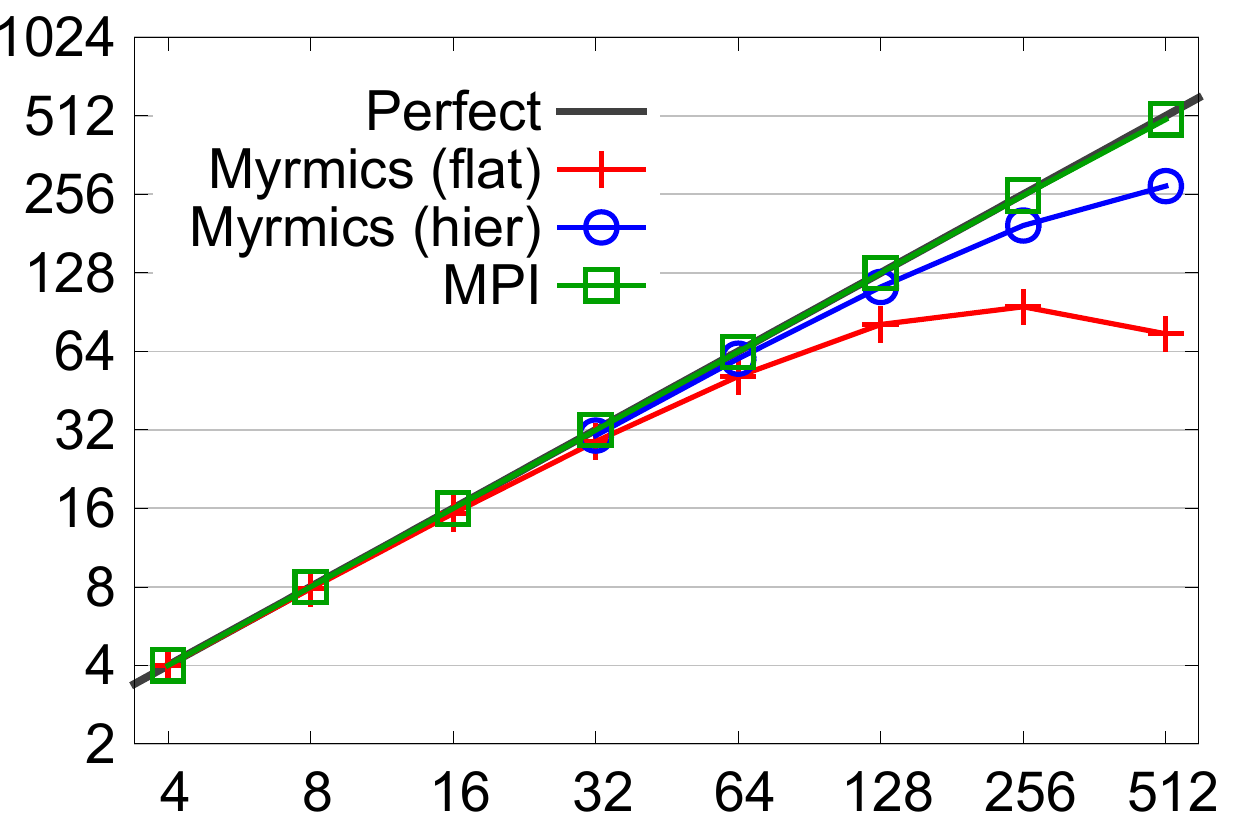}
    \label{fig:kmeans_strong}
  }
  \hfill
  \subfloat[Matrix Mult. (strong)]{
    \includegraphics[width=0.315\textwidth]{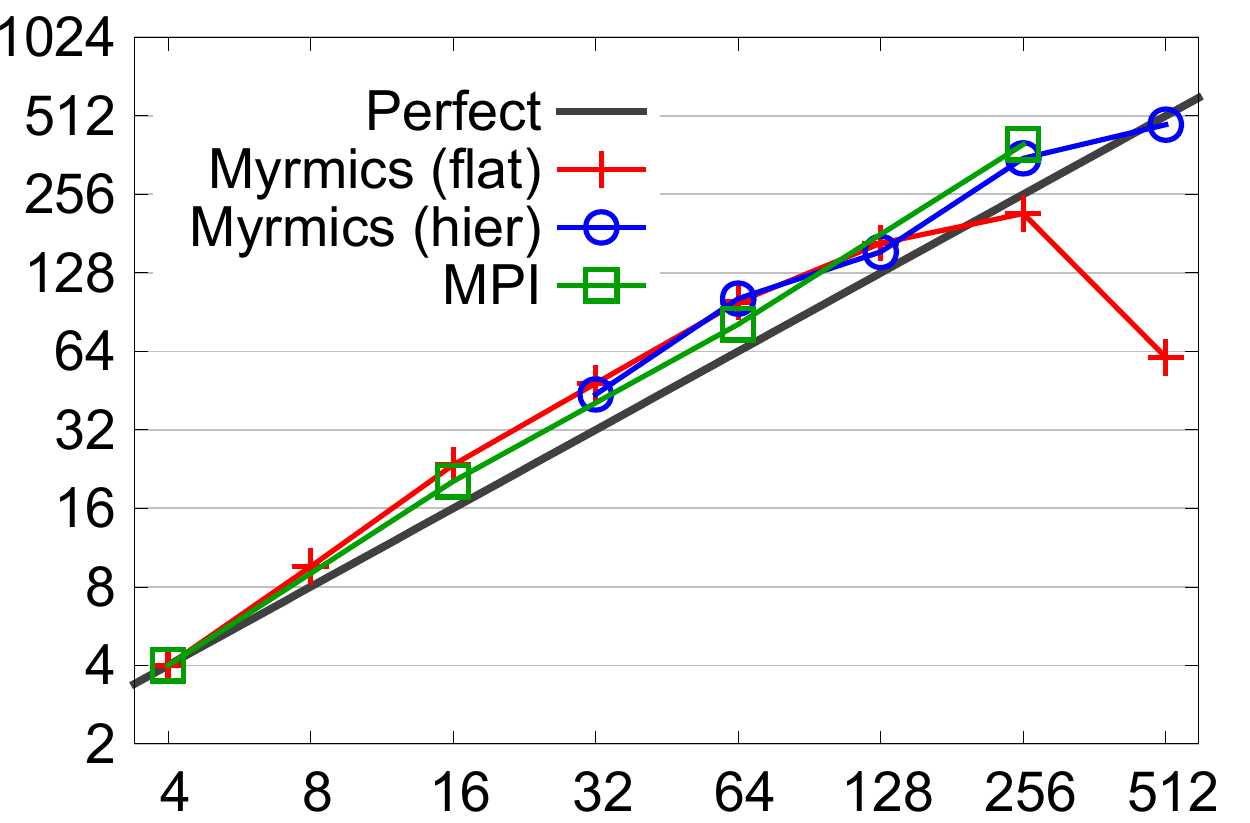}
    \label{fig:matrix_strong}
  }
  \hfill
  \subfloat[Barnes-Hut (strong)]{
    \includegraphics[width=0.315\textwidth]{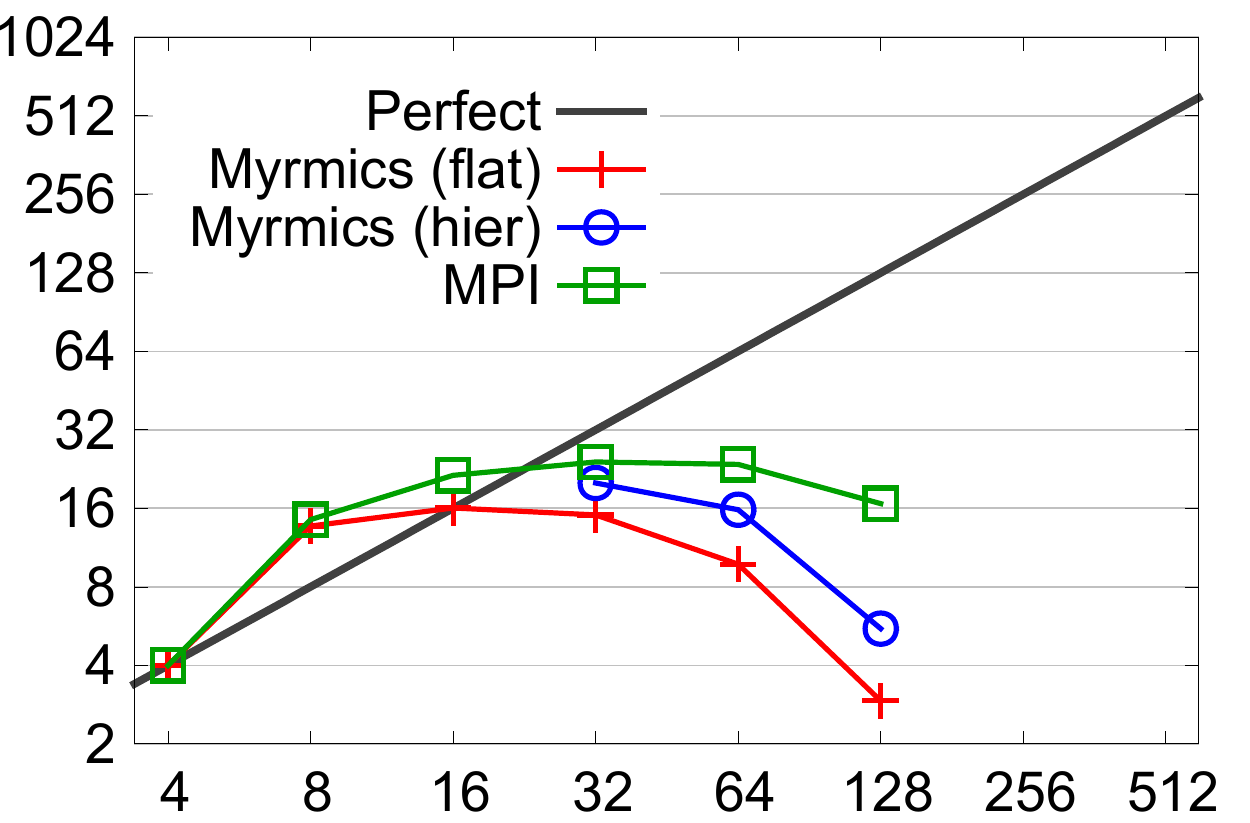}
    \label{fig:barnes_strong}
  }

  \subfloat[Jacobi Iteration (weak)]{
    \includegraphics[width=0.315\textwidth]{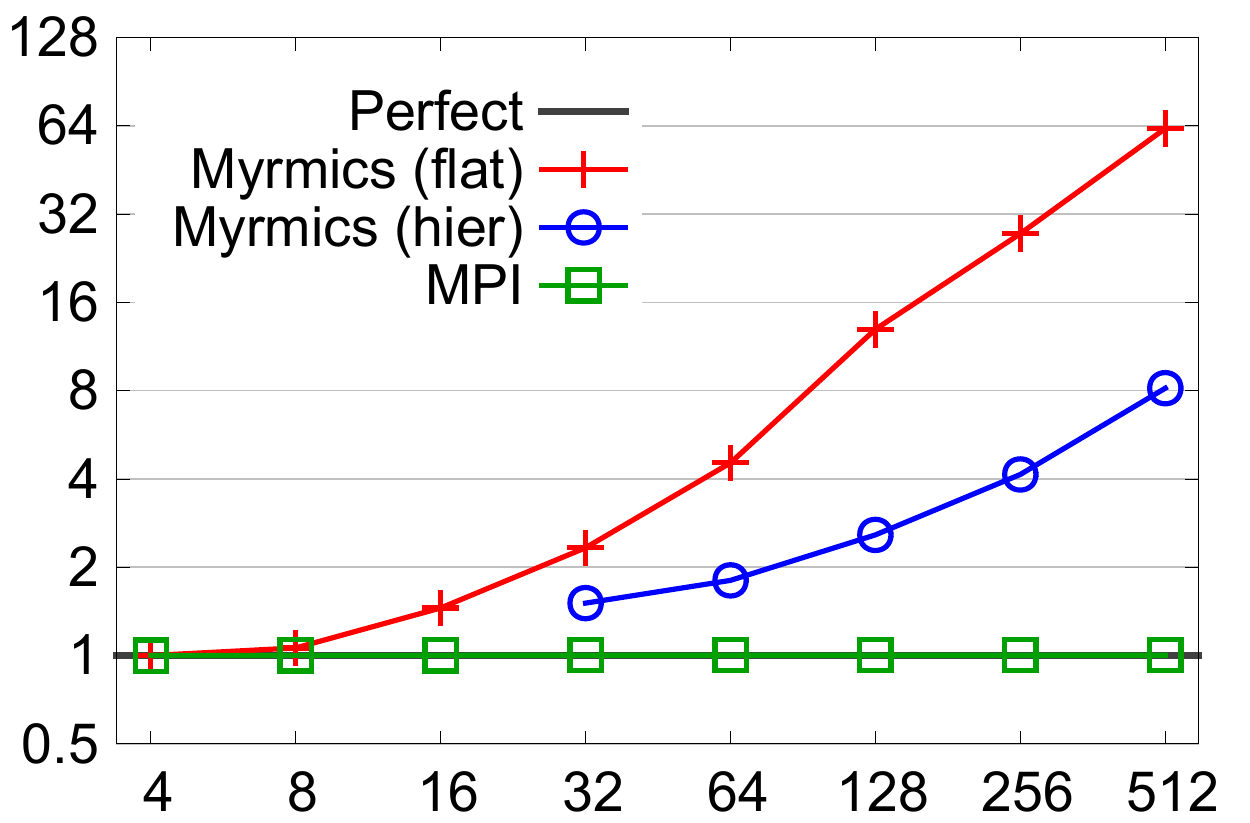}
    \label{fig:jacobi_weak}
  }
  \hfill
  \subfloat[Raytracing (weak)]{
    \includegraphics[width=0.315\textwidth]{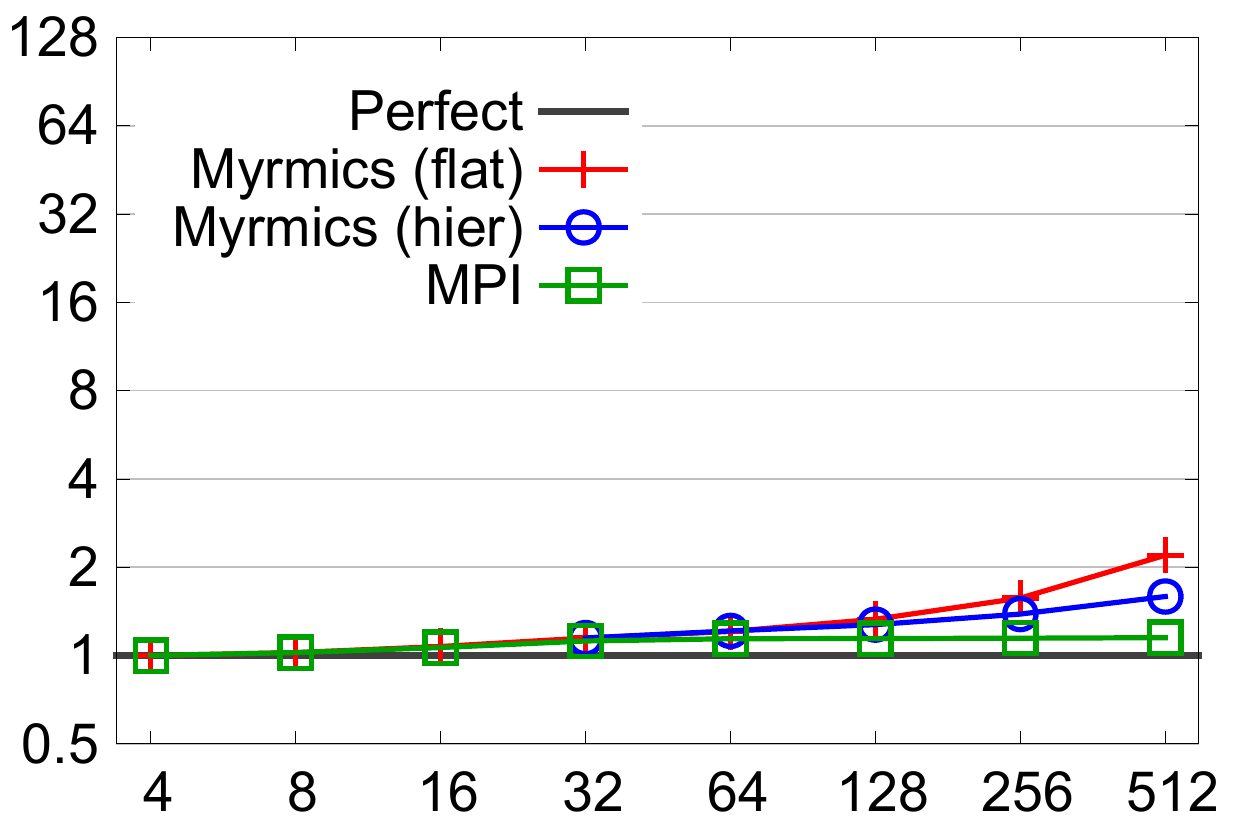}
    \label{fig:cray_weak}
  }
  \hfill
  \subfloat[Bitonic Sort (weak)]{
    \includegraphics[width=0.315\textwidth]{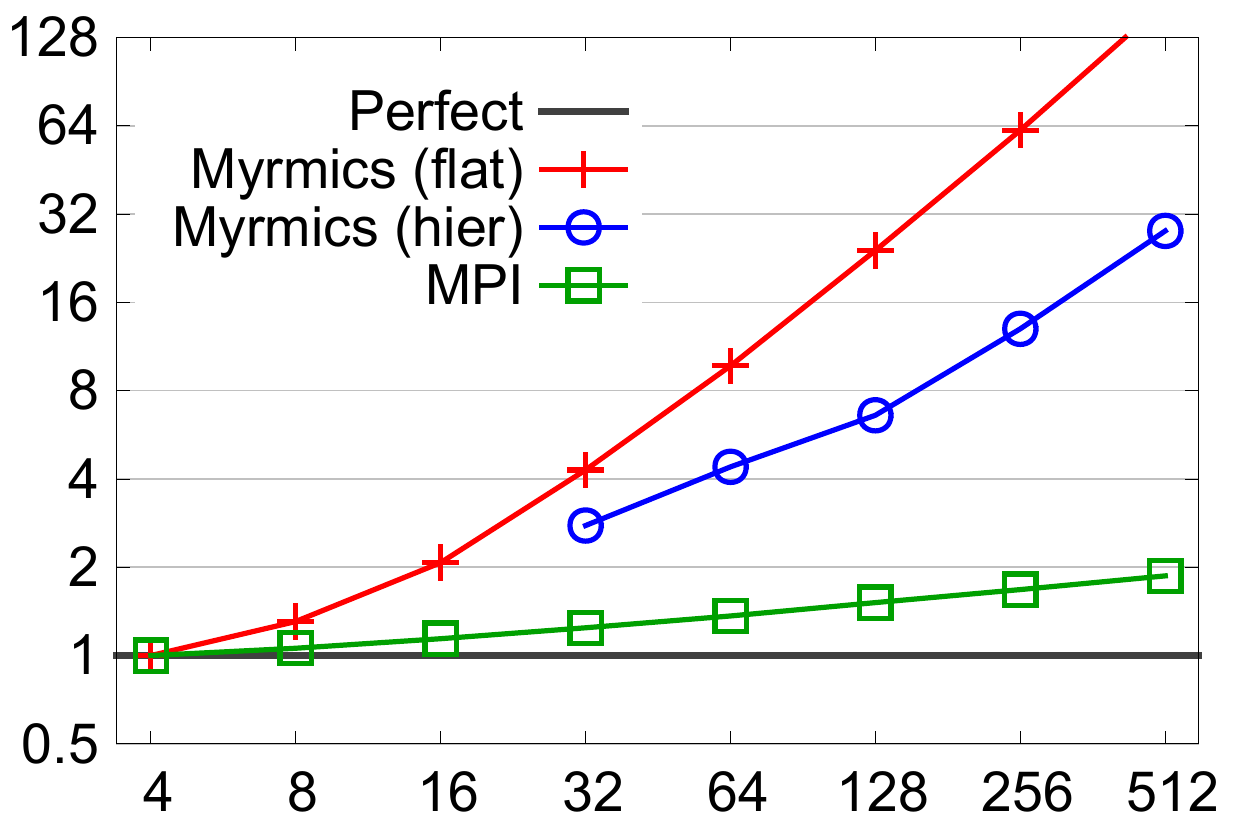}
    \label{fig:bitonic_weak}
  }

  \subfloat[K-Means Clustering (weak)]{
    \includegraphics[width=0.315\textwidth]{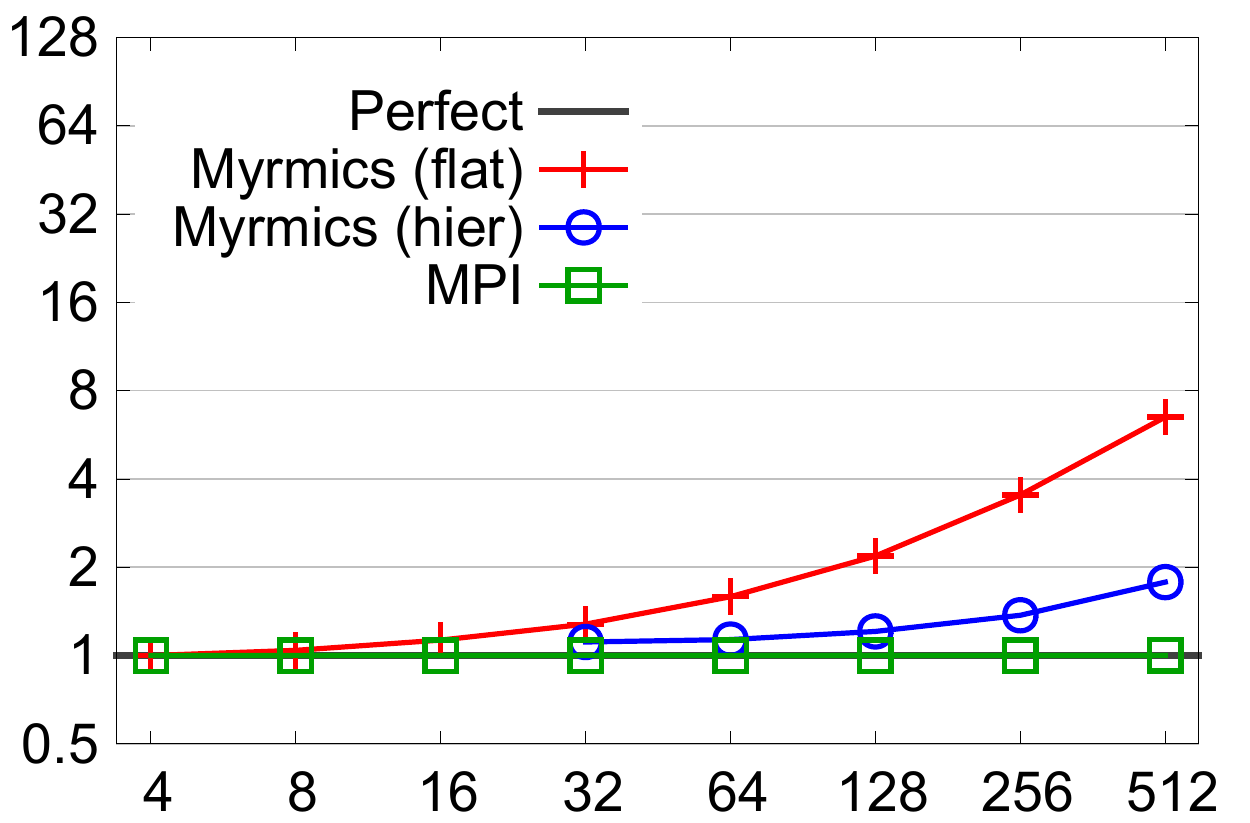}
    \label{fig:kmeans_weak}
  }
  \hfill
  \subfloat[Matrix Mult. (weak)]{
    \includegraphics[width=0.315\textwidth]{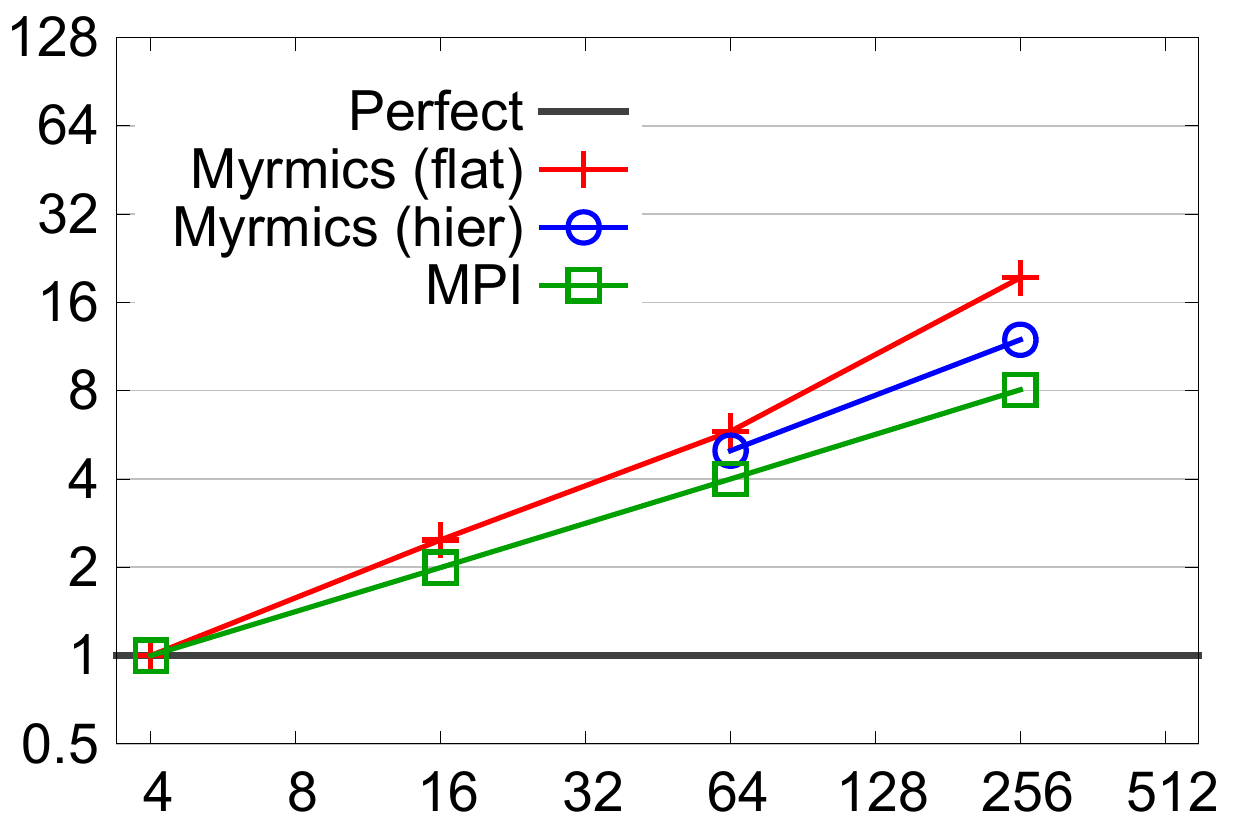}
    \label{fig:matrix_weak}
  }
  \hfill
  \subfloat[Barnes-Hut (weak)]{
    \includegraphics[width=0.315\textwidth]{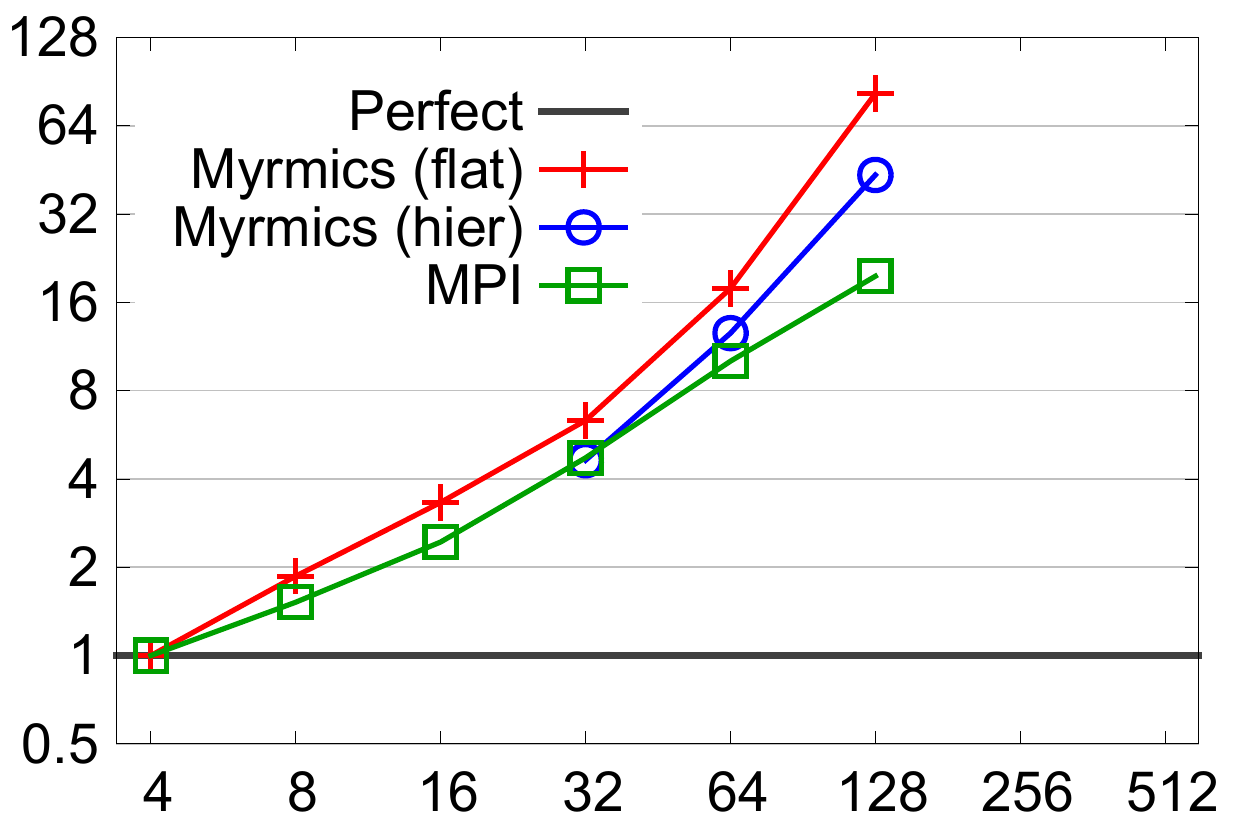}
    \label{fig:barnes_weak}
  }

  \caption{
    Myrmics and MPI scaling. Strong scaling results in (a)--(f),
    weak scaling in (g)--(l). In all graphs, X axis measures the
    number of worker cores (Myrmics) or total cores (MPI).
    Scheduler cores for Myrmics are as follows: 1 core for flat
    scheduling, or 1 top-level scheduler plus L leaf schedulers
    for hierarchical configurations, where L=2 for 32 workers,
    L=4 for 64 workers and L=7 for 128, 256 or 512 workers. Y
    axis measures speedup for strong scaling or slowdown for
    weak scaling, normalized to single-worker performance.
  }
  \label{fig:scaling}
\end{figure*}

We use five benchmark kernels to test diverse kinds of parallel
communication behaviors. 
\textit{Jacobi Iteration} is a subset of a linear algebra iterative
solver.  Its scaling results are shown in
Figs.~\ref{fig:jacobi_strong} and~\ref{fig:jacobi_weak}.  A
table of values with a fixed border is split into multiple
workers, where each worker takes a succession of rows. In each
loop repetition, every table element is replaced by the average
of its four neighbors (north, east, south, west).  Jacobi
exhibits a nearest-neighbor communication pattern, because across
loop boundaries each worker receives the top and bottom rows of
its neighbors. We use regions to split the table into groups of
rows.
In \textit{Raytracing}
(Figs.~\ref{fig:cray_strong}, \ref{fig:cray_weak}), a description
of a scene geometry (objects, lights, camera) is made available
to all workers. Each worker renders a part of a picture frame, by
computing how light rays from the camera to the frame pixels
interact with the scene objects and lights. This kernel is
embarrassingly parallel, since apart from loading the scene
description each worker computes its own frame parts in
isolation. We use regions to split the frame into groups of pixel
lines. In \textit{Bitonic Sort}
(Figs.~\ref{fig:bitonic_strong}, \ref{fig:bitonic_weak}), each
worker begins with a part of the data to be sorted, and sorts
this part. Afterwards, in a number of stages equal to the squared
binary logarithm of the number of cores, workers exchange data
and merge-sort their local buffers with the incoming ones.
Bitonic Sort exhibits butterfly-like communication among workers
in the data exchange phase. The data to be sorted are divided
into coarse regions when the algorithm initializes.
\textit{K-Means Clustering}
(Figs.~\ref{fig:kmeans_strong}, \ref{fig:kmeans_weak})
heuristically groups a big number of 3D objects into a few
clusters based on the objects proximity. Beginning with a random
cluster assignment, in each iteration the workers assign their
share of objects into the clusters. In the end of each iteration,
clusters are recomputed to be at the center of grouped objects.
K-Means Clustering features parallel reductions and broadcasts.
We use two kinds of regions in this kernel. First, the objects to
be clustered are divided into a number of regions. Second, we
employ a few regions to hold the temporary buffers during the
reductions at the end of each loop. 
\textit{Matrix Multiplication}
(Figs.~\ref{fig:matrix_strong}, \ref{fig:matrix_weak})
multiplies two dense arrays. Each worker has a part of the two
source arrays and of the destination array. During each phase, a
worker adds partial multiplication results to its destination
array by doing a matrix multiplication of smaller parts of the
two source arrays. This kernel exhibits communication bursts, as
parts of the source arrays temporarily become hot spots which are
shared by multiple workers for a computation phase. All three
matrices are split into a number of regions, each containing a 2D
piece of a matrix.
Finally, we also evaluate \textit{Barnes-Hut}, an
application that uses pointer-based data structures and exhibits
irregular parallelism (Figs.~\ref{fig:barnes_strong},
\ref{fig:barnes_weak}). Barnes-Hut solves efficiently an N-body
problem by grouping far-away collections of bodies into single
bodies. The application makes heavy use of dynamically allocated
trees, which are built and destroyed in each step of the
algorithm. Each computation task allocates a tree for its local
bodies; this tree belongs to a new region, which is created for
the loop repetition and destroyed when the repetition ends.
Bodies are allocated in these regions to create the Barnes-Hut
octrees. To compute the gravitational forces, tasks are created
to operate on two regions, each containing an octree of a part of
the 3D space. We further describe the parallelization of this
application, as well as how regions can be used effectively to
program irregular applications, in our previous
work~\cite{myrmics_ismm12}.

We observe that MPI benchmarks
scale almost perfectly (green/square lines in all figures). This
is expected, both because we employ well-known parallelization
methods and also because we a lightweight MPI library
implementation which runs on an emulated architecture of a
single-chip manycore CPU with a very efficient network-on-chip.
We can therefore depend on the MPI benchmarks to provide a solid
baseline for comparing the Myrmics performance on the same
architecture. Super-optimal scaling is present in some strong
scaling cases (Figs.~\ref{fig:bitonic_strong},
\ref{fig:matrix_strong}) where the per-worker task dataset fits
entirely in the caches. Under-optimal weak scaling
(Figs.~\ref{fig:bitonic_weak}, \ref{fig:matrix_weak}) is expected, as
these algorithms have non-linear complexity when adding more
workers. The Matrix Multiplication graphs have fewer data points,
because the algorithm depends on the number of cores being a
power of 4. Barnes-Hut does not scale well,
because it involves many and communication-intensive steps, such
as load-balancing exchanges, all-to-all communication and phases
with idle workers. We do not include numbers for 256 and 512
cores due to memory constraints, but the scaling already degrades
after 64 cores.

We next focus on how Myrmics scales using a
single scheduler (red/cross lines). Note that the single
scheduler performs well up to a certain number of workers,
depending on the benchmark. As we use a minimum task size of 1 M
cycles, we expect the turning point to be 64 workers. This is
confirmed for benchmarks that have bigger
communication/computation ratio, such as Jacobi and Bitonic,
while others are less affected.

We verify our core hypothesis on hierarchical scheduling by
observing how Myrmics scales when using a two-level hierarchy of
schedulers (blue/circle lines). In all benchmarks, the
multiple-scheduler setup outperforms the single scheduler. 
This happens because the schedulers manage to share the load of
the workers efficiently, each low-level scheduler being directly
responsible for a subset of the total workers.
The Myrmics benchmarks code is
written in a hierarchical way to support this. The application
decomposes the dataset into a number of regions. It then
specifies a few number of high-level tasks that operate on whole
regions (\textit{e.g.} to perform one loop iteration, or do a
reduction on whole regions). These tasks spawn children tasks
that operate on some of the region objects. Myrmics assigns the
the few high-level tasks to the top-level schedulers and the many
low-level tasks to the low-level schedulers. Thus, the
application run is mostly contained into multiple local ``domains'',
each consisting of a low-level scheduler and its workers.
Messages and DMA transfers are localized and the application can
scale much better than with the single scheduler setup. 

As explained in Fig.~\ref{fig:scaling}, in runs up to 128
worker cores we maintain that each low-level scheduler is
responsible for up to 16--18 workers. Our experiments show that
this is an optimal point for scheduler-to-worker ratio. This is
less than the 64 workers we computed with the microbenchmark in
section~\ref{sec:overhead}, but it is reasonable for real-life
benchmarks that have multiple arguments per task, with multiple
dependencies. Since our hardware setup is limited to 8 total
Cortex-A9 cores, the 256- and 512-worker cores configuration is
sub-optimal, with each low-level scheduler handling up to 37 and
74 workers respectively. We believe this to be the main reason
that these data points show a degradation in scaling.


The execution time for Myrmics benchmarks is usually higher than
the respective MPI ones. The numbers vary greatly, depending
mostly on whether the Myrmics version scales well on the selected
core count we measure or not. We find that a typical overhead for
data points that scale well is in the range of 10\%--30\%. This
overhead represents the time the runtime needs to perform all the
auto-parallelization work. There are cases where this overhead
can be minimized, \textit{e.g.} by over-decomposing a very
parallel problem into many tasks; the runtime can complete its
work in the background using the scheduler cores, while all
workers are kept busy. However, there are also cases that this
cannot be avoided, such as when reductions must be done across
loop boundaries. Furthermore, our experiments analysis indicates
that Myrmics automatic data placement algorithms work quite well.
Myrmics overhead compared to MPI is attributed to the dependency
analysis and scheduling, and is not caused by any excessive
remote data transfers.

\subsection{Qualitative Analysis}

\begin{figure*}
  \subfloat[Bitonic Sort]{
    \includegraphics[width=0.315\textwidth]{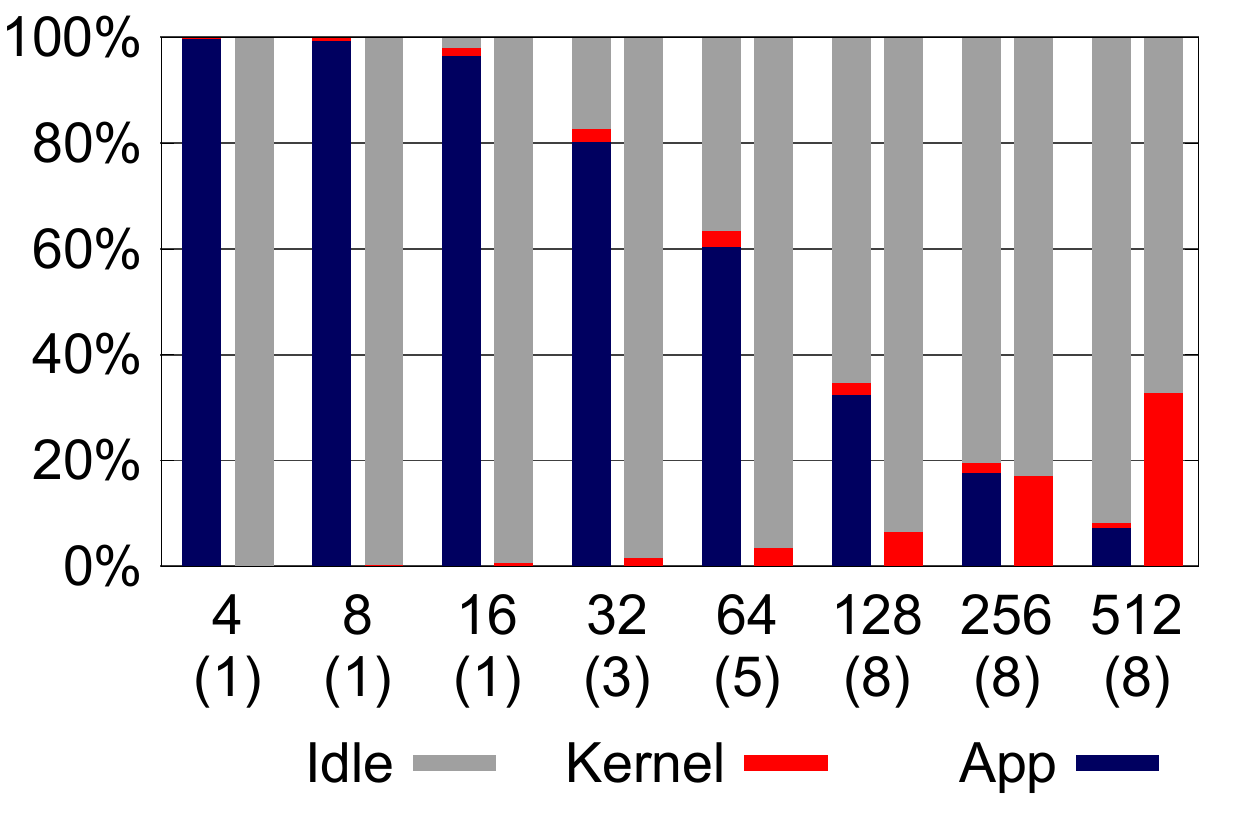}
    \label{fig:bitonic_breakdown}
  }
  \hfill
  \subfloat[K-Means Clustering]{
    \includegraphics[width=0.315\textwidth]{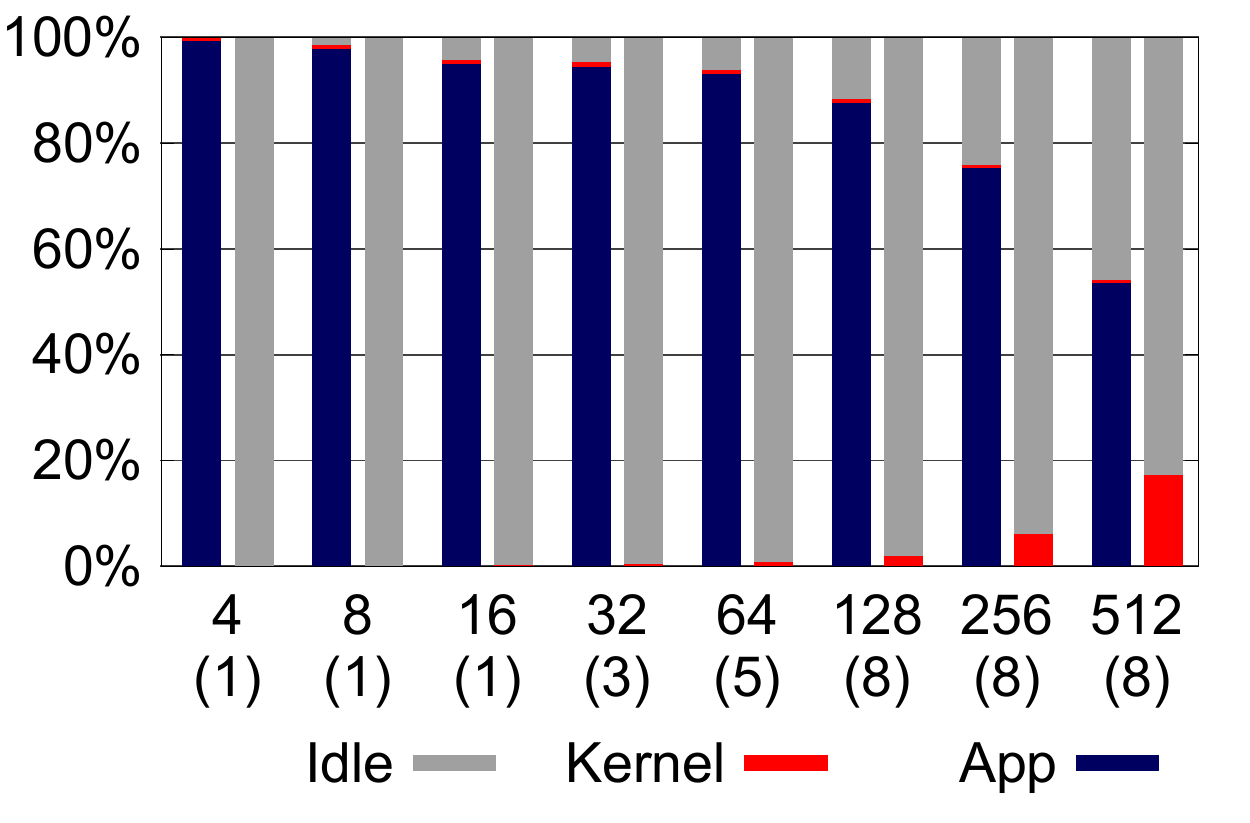}
    \label{fig:kmeans_breakdown}
  }
  \hfill
  \subfloat[Raytracing]{
    \includegraphics[width=0.315\textwidth]{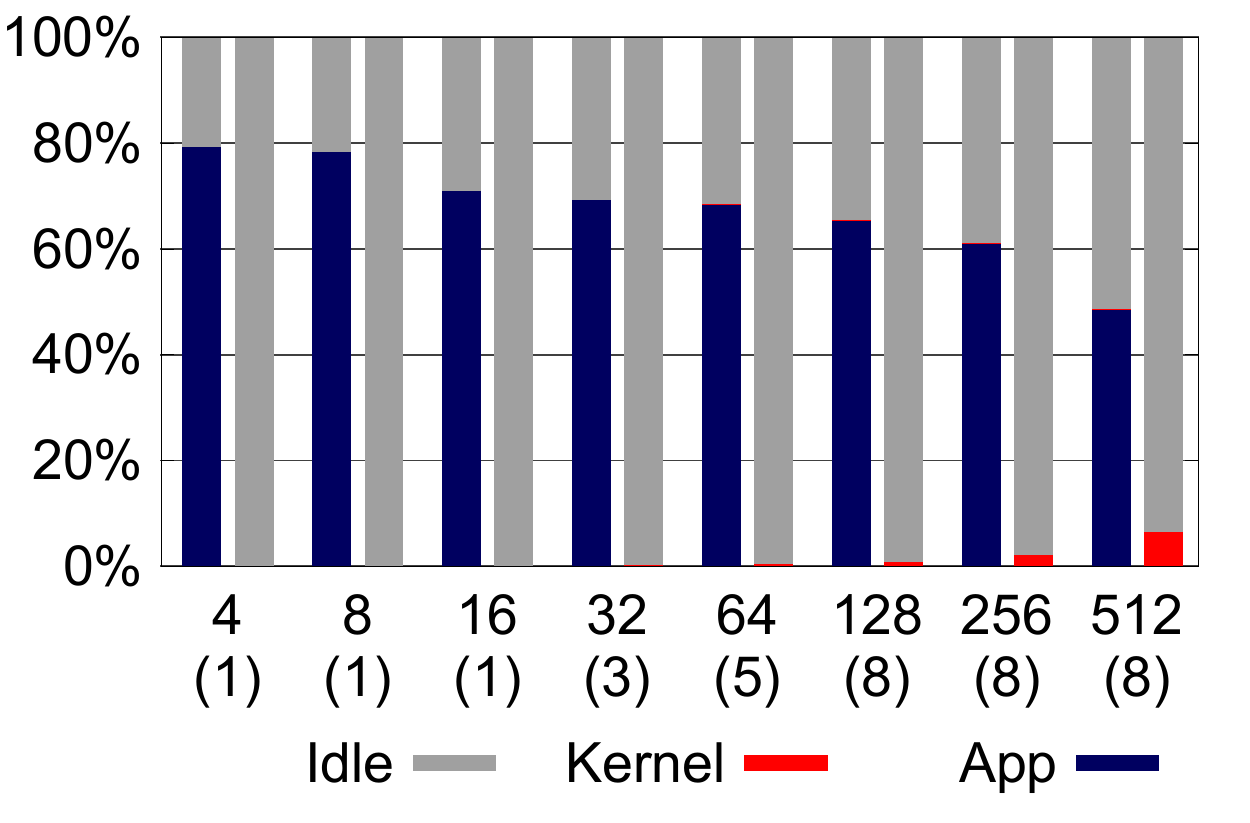}
    \label{fig:cray_breakdown}
  }

  \caption{
    Time breakdown. X axis shows the number of worker cores and
    below the number of scheduler cores (in parentheses). Y axis
    measures percentages, based on the total execution time. The
    left bar in a pair indicates where a worker core spent its
    time. The right bar indicates the same for a scheduler core.
    The bars are averaged per worker or scheduler core
    respectively.
  }
  \label{fig:breakdown}
\end{figure*}

To further understand how scheduler and worker cores in Myrmics
perform, we select three of the kernels and study their strong
scaling executions in more depth. We choose the worst-performing
kernel (Bitonic Sort), a medium case (K-Means) and the
best-performing one (Raytracing). We first gather statistics
about the breakdown of time inside the schedulers and workers.
Results are shown in Fig.~\ref{fig:breakdown}. In Bitonic Sort
(Fig.~\ref{fig:bitonic_breakdown}), this analysis reveals the
reason it scales poorly. In high core counts, most workers (left
bars) spend their time being idle (gray/light) instead of running
application tasks (blue/dark), while the schedulers' load (right
bars, red/medium) increases. Depending on the phase of the
bitonic sorting, the benchmark may spawn a big number of tasks
and the schedulers are not fast enough to handle it. However, if
we decrease the dataset decomposition to spawn less tasks, then
there are other application phases where the number of tasks is
too small and the performance is degraded due to lack of
parallelism. A general observation from our experiments is that
when a scheduler is over 10\% busy, it does not process
requests fast enough to be considered responsive. In the
512-worker Bitonic Sort case, the average scheduler load is 33\%
and the system is significantly slowed down. Our analysis
indicates that scheduler responsiveness is the main reason the
system slows down. Overhead due to DMA transfers is negligible,
which indicates that good data locality is achieved.

K-Means Clustering (Fig.~\ref{fig:kmeans_breakdown}) results show
that the workers are kept busy executing tasks for higher core
counts than in the Bitonic Sort case. This is a more typical
behavior, since this benchmark spawns an equal number of tasks
per computation step. Up to 128 workers, workers are executing
tasks for 88\% of their time while schedulers are busy 2\% of
their time. In the 512-worker case, these numbers become 53\% and
17\% respectively and the performance begins to suffer. In
Raytracing (Fig.~\ref{fig:cray_breakdown}) we see an even more
ideal situation.  The total number of tasks is small compared to
the other two benchmarks and the work is embarrassingly parallel.
We observe that the scheduler load is at the worst case 6\%, and
indeed the benchmark scales well. The workers are busy between
79\% of their time at best (4 workers) and 48\% at worst (512
workers). The fact that the workers are not fully busy at low
core counts is explained by the way the benchmark decomposes the
dataset: it assigns chunks of work equal to the picture lines
divided by the available workers. Thus, the working set for each
core has the same amount of picture lines which does not
necessarily imply the same amount of work, as the latter depends
on the complexity of the scene ---some picture lines will be in
the path of more scene objects than others.

\begin{figure*}
  \subfloat[Bitonic Sort]{
    \includegraphics[width=0.315\textwidth]{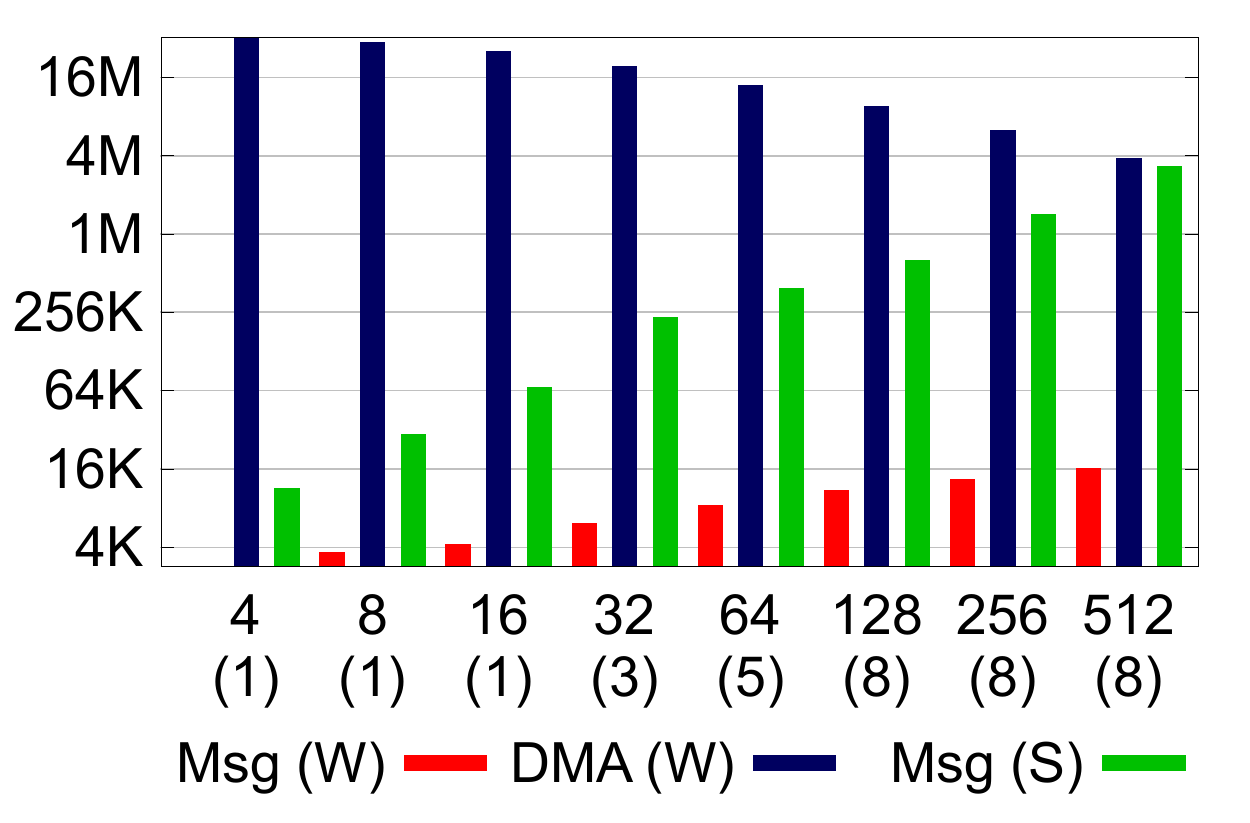}
    \label{fig:bitonic_traffic}
  }
  \hfill
  \subfloat[K-Means Clustering]{
    \includegraphics[width=0.315\textwidth]{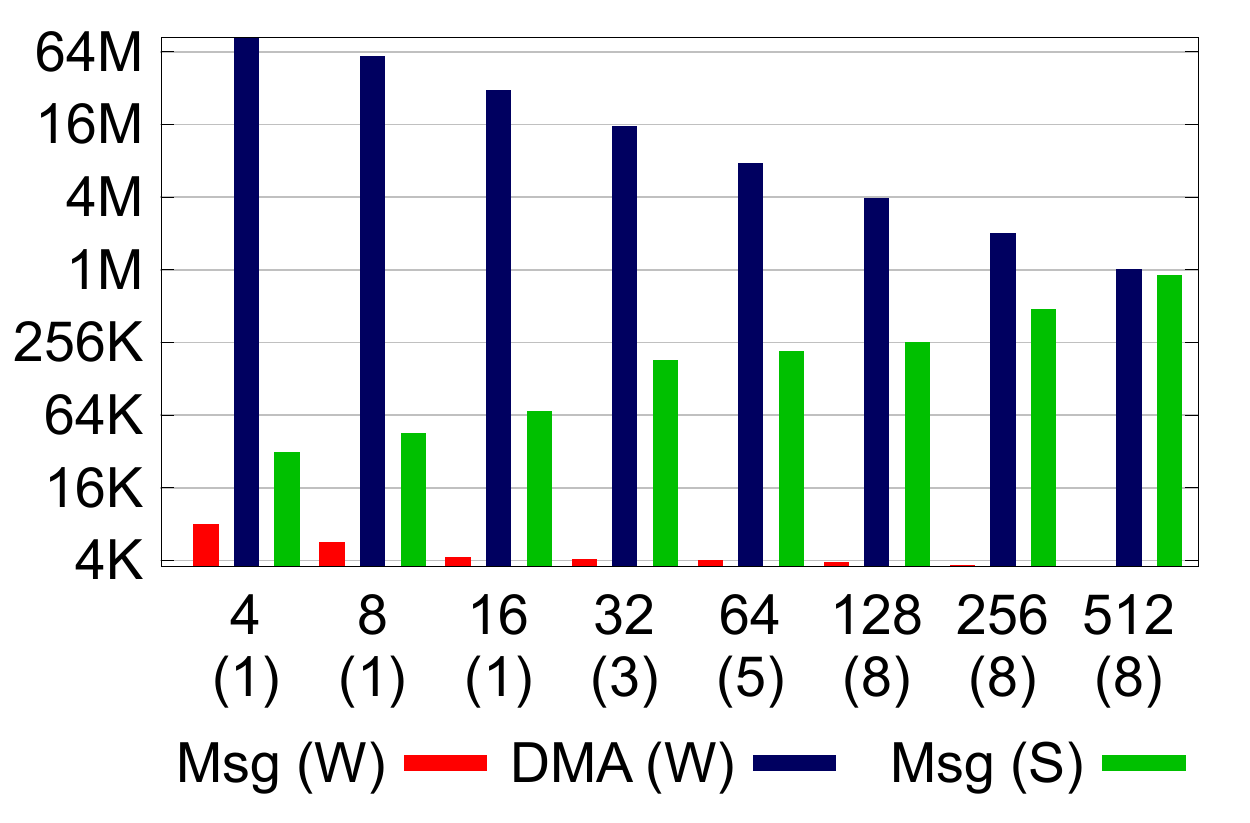}
    \label{fig:kmeans_traffic}
  }
  \hfill
  \subfloat[Raytracing]{
    \includegraphics[width=0.315\textwidth]{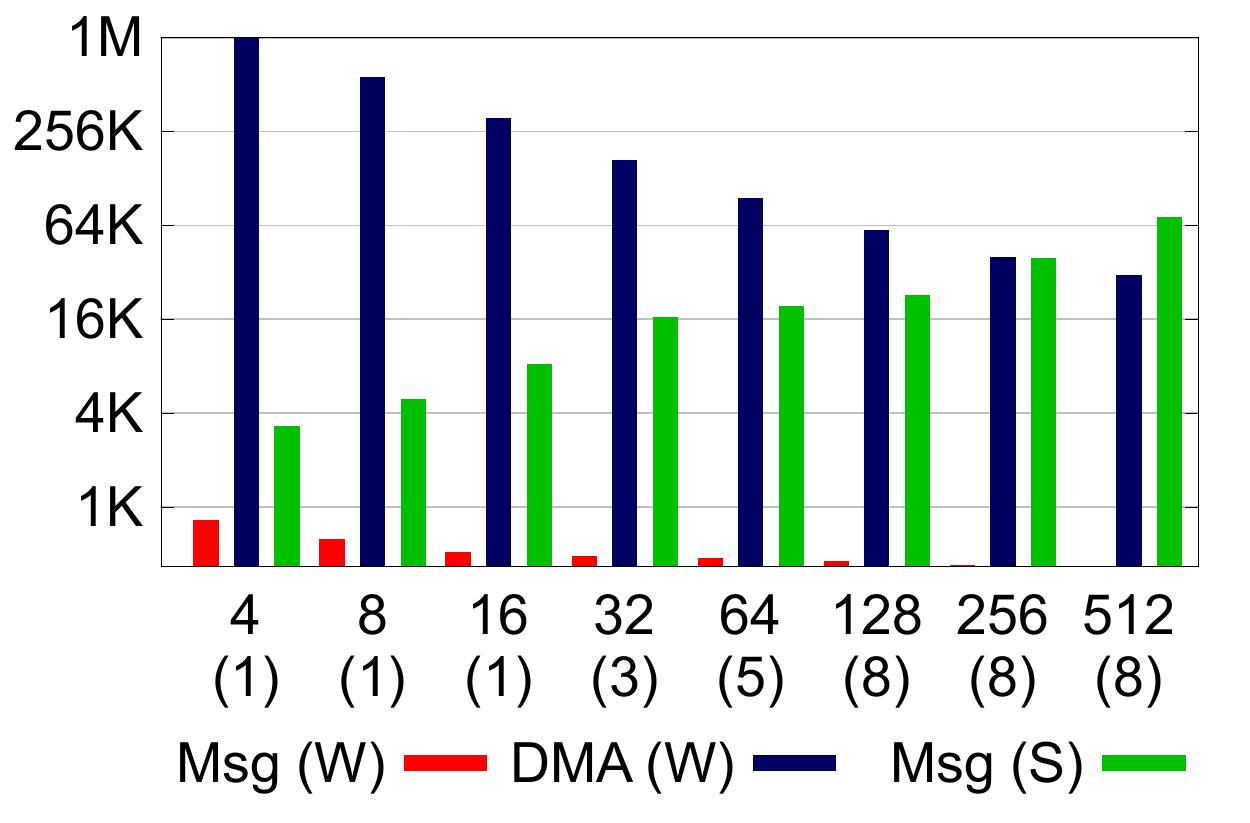}
    \label{fig:cray_traffic}
  }

  \caption{
    Traffic analysis. X axis shows the number of worker cores and
    below the number of scheduler cores (in parentheses). Y axis
    is logarithmic and measures core communication in bytes. The
    first bar in a triplet (red/halftone) counts the worker
    message volume, the second bar (blue/dark) counts the worker
    DMA transfer volume and the third bar (green/light) counts
    the scheduler message volume. The bars are averaged per
    worker or scheduler core respectively.
  }
  \label{fig:traffic}
\end{figure*}

Fig.~\ref{fig:traffic} shows our second set of qualitative
measurements for the same benchmarks. When an application scales
gracefully, we expect the worker-scheduler message traffic and
worker DMA traffic to decrease (per worker core), as each worker
runs a smaller piece of the problem. We also expect the scheduler
message traffic to increase, as the schedulers collectively spawn
more tasks for more workers. Note the pathological case of the bad
Bitonic Sort behavior for high core counts. 
The average per-scheduler message-based
communication (green/light bars) rises much more rapidly in
Bitonic Sort than the other benchmarks, and reaches a very high
peak at 512 workers (4 MB, instead of 256 KB and 64 KB). This is
indicative of too many spawned tasks, which is also reflected in
the average per-worker messages (red/medium bars). In Bitonic
Sort the worker-scheduler communication increases at higher core
counts; in the other two benchmarks it slightly decreases, as
expected. For strong scaling benchmarks with a variable task
size, we expect a worker core to execute roughly the same number
of tasks at higher core counts, each task dealing with a smaller
part of the total dataset. The worker message traffic in Bitonic
Sort indicates that workers execute more tasks as the benchmark
scales. However, in all three cases the DMA transfer
communication per worker (blue/dark bars) decreases. This is an
artifact of the tasks being smaller.

\subsection{Locality \textit{vs.} Load-Balancing}
\label{sec:policy_eval}

\begin{figure*}
  \subfloat[Matrix Mult. (32 wrk)]{
    \includegraphics[width=0.315\textwidth]{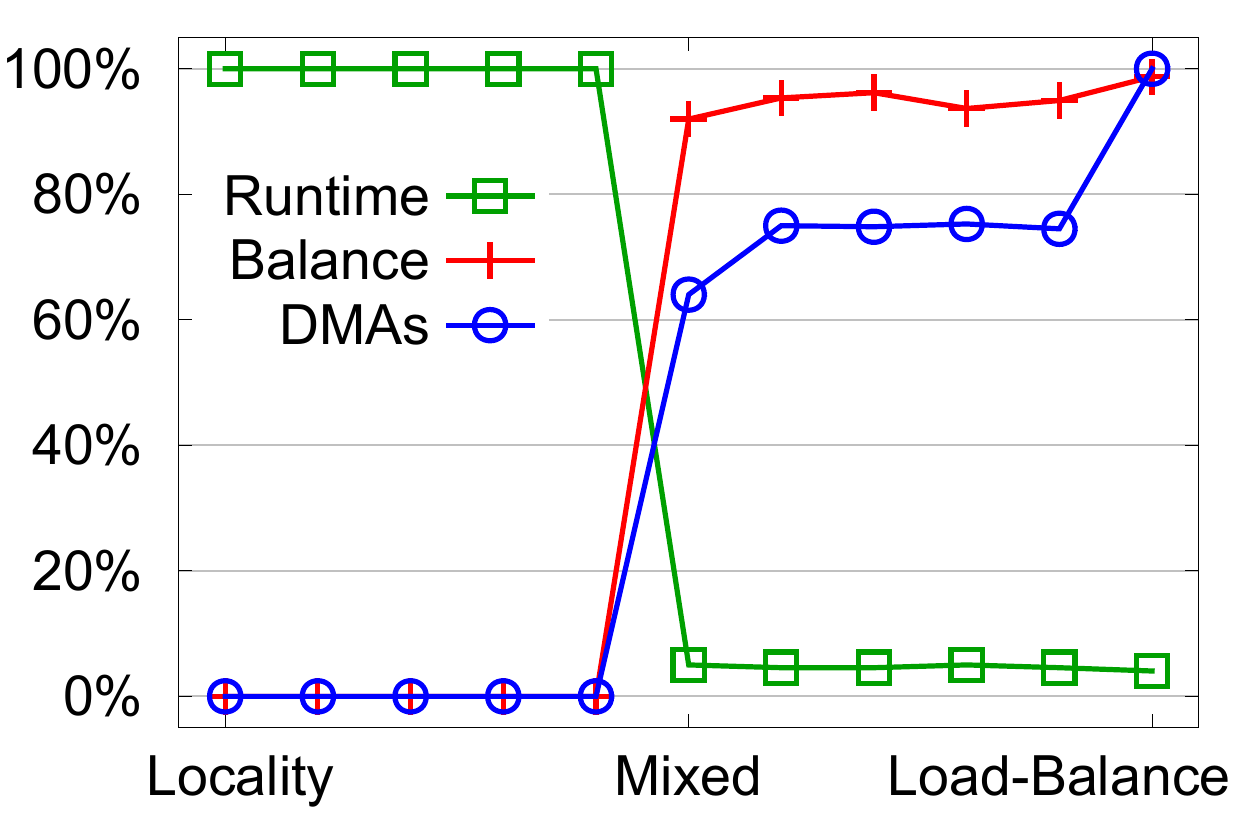}
    \label{fig:matrix_locality}
  }
  \hfill
  \subfloat[Jacobi Iteration (128 wrk)]{
    \includegraphics[width=0.315\textwidth]{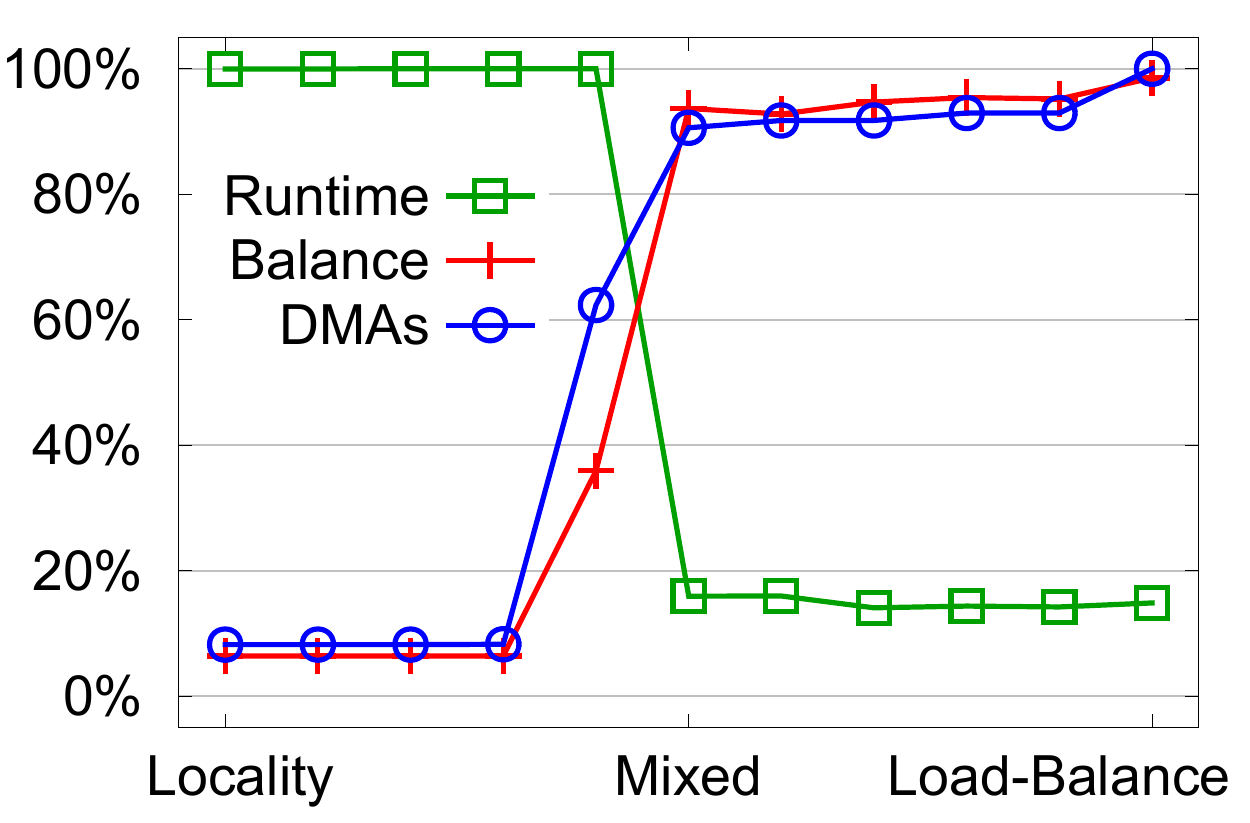}
    \label{fig:jacobi_locality}
  }
  \hfill
  \subfloat[K-Means Clustering (512 wrk)]{
    \includegraphics[width=0.315\textwidth]{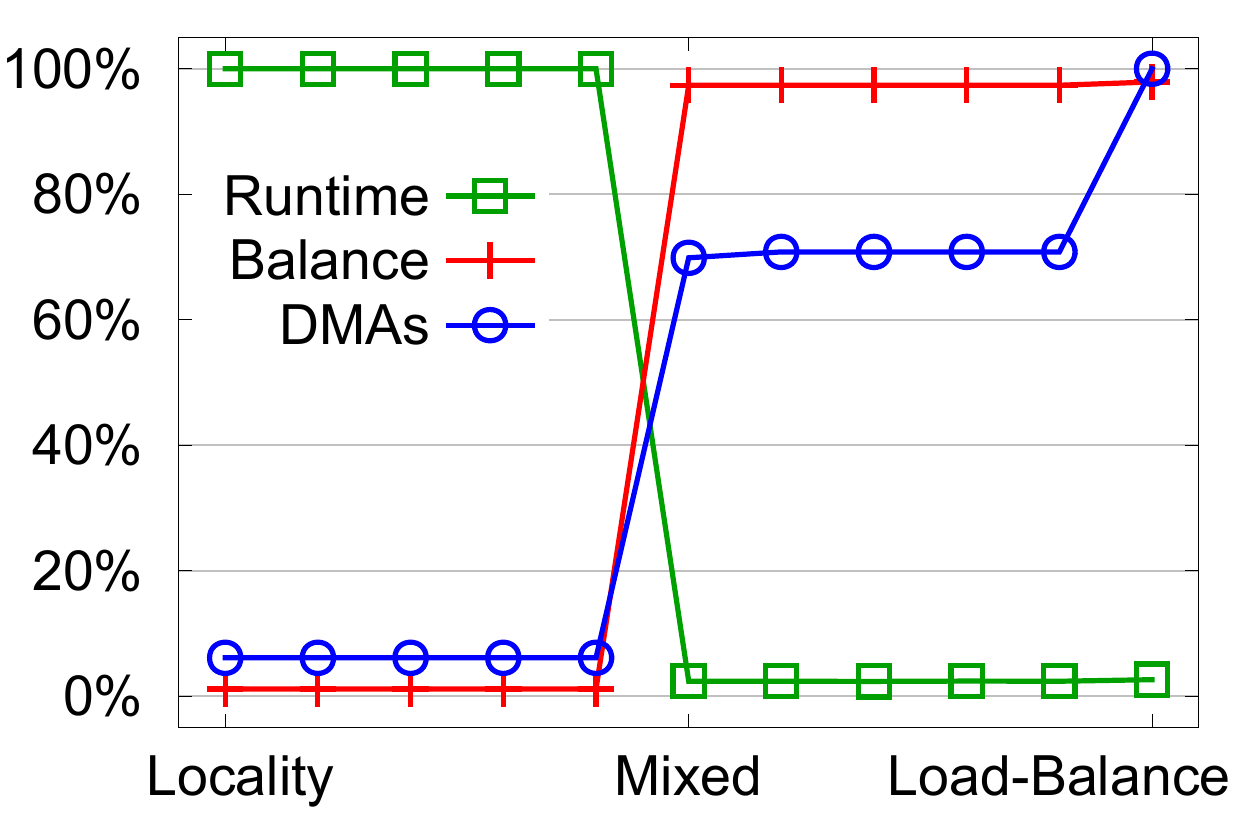}
    \label{fig:kmeans_locality}
  }

  \caption{
    Effect of load-balancing vs. locality scheduling criteria. 
    The X axis shows how much we favor the locality scheduling
    score (left X values, $p$=100) to the load-balancing score (right X
    values, $p$=0). The Y axis shows how this choice impacts the
    application running time, the system-wide load balance and
    the total DMA traffic. Y values are normalized to the
    maximum for this experiment and measured in percentages. We
    measure system-wide load balance as the average deviation of
    the tasks each worker core runs \textit{vs.} the optimal
    number of tasks it should run: 100\% balance means each
    worker runs exactly (total tasks)/(number of workers) and 0\%
    balance means one worker runs (total tasks) and all others
    are idle.
  }
  \label{fig:locality}
\end{figure*}

As explained in section~\ref{sec:scheduling}, when a task is
dependency-free the schedulers cooperate to progressively
schedule it down the hierarchy. Two scores are computed, one
favoring subtrees of workers where the arguments the task needs
were last produced (a ``locality'' score $L$), and one favoring
subtrees of workers that are idle or less busy than others (a
``load-balance'' score $B$). The total score is $T = pL + (100 -
p)B$, where $p$ is a policy bias percentage value. We run a
series of experiments that sweeps $p$, to affect the relative
weights of $L$ and $B$. The results are shown in
Fig.~\ref{fig:locality}. We use the Matrix Multiplication kernel
with flat scheduling and 32 workers
(Fig.~\ref{fig:matrix_locality}), the Jacobi Iteration with
hierarchical scheduling and 128 workers
(Fig.~\ref{fig:jacobi_locality}) and the K-Means Clustering with
hierarchical scheduling and 512 workers
(Fig.~\ref{fig:kmeans_locality}).

As expected, these two scores are conflicting. Perfect locality is
maintained only when using a single worker (with a single scheduler),
or a single worker sub-tree 
(hierarchical). This minimizes the DMA transfers
communication, but on the other hand causes the application
running time to suffer, as only one or a few workers are busy.
Taking into account the load-balancing score causes more communication
but also improves the application
running time. If we only use the load-balancing score (far
right point in the graphs), there is an increased communication
volume, as the schedulers do not optimize at all for locality.
Although not very apparent in the graphs, there is a noticeable
performance degradation in this case ---\textit{e.g.} in
K-Means, the load-balance-only point is 10\% worse in running
time \textit{vs.} the previous one. We found a good
trade-off between running time and communication volume lies in
the range of assigning a 0.7--0.9 load-balance weight and a
0.3--0.1 locality weight respectively.

\subsection{Deeper Hierarchies}
\label{sec:deeper}

\begin{figure}
\centering
  \subfloat[]{
    \includegraphics[width=0.8\columnwidth]{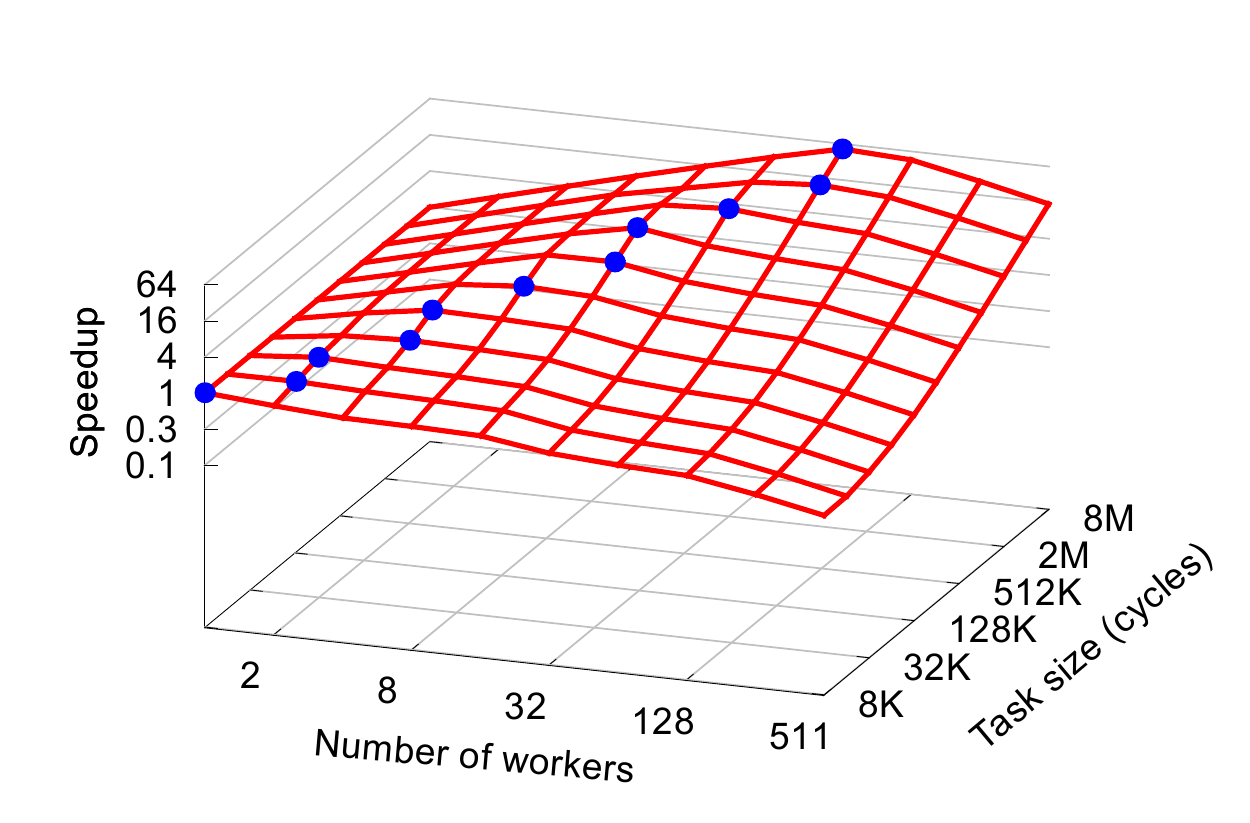}
    \label{fig:task_synthetic_mb}
  }
  \hfill
  \subfloat[]{
    \includegraphics[width=0.7\columnwidth]{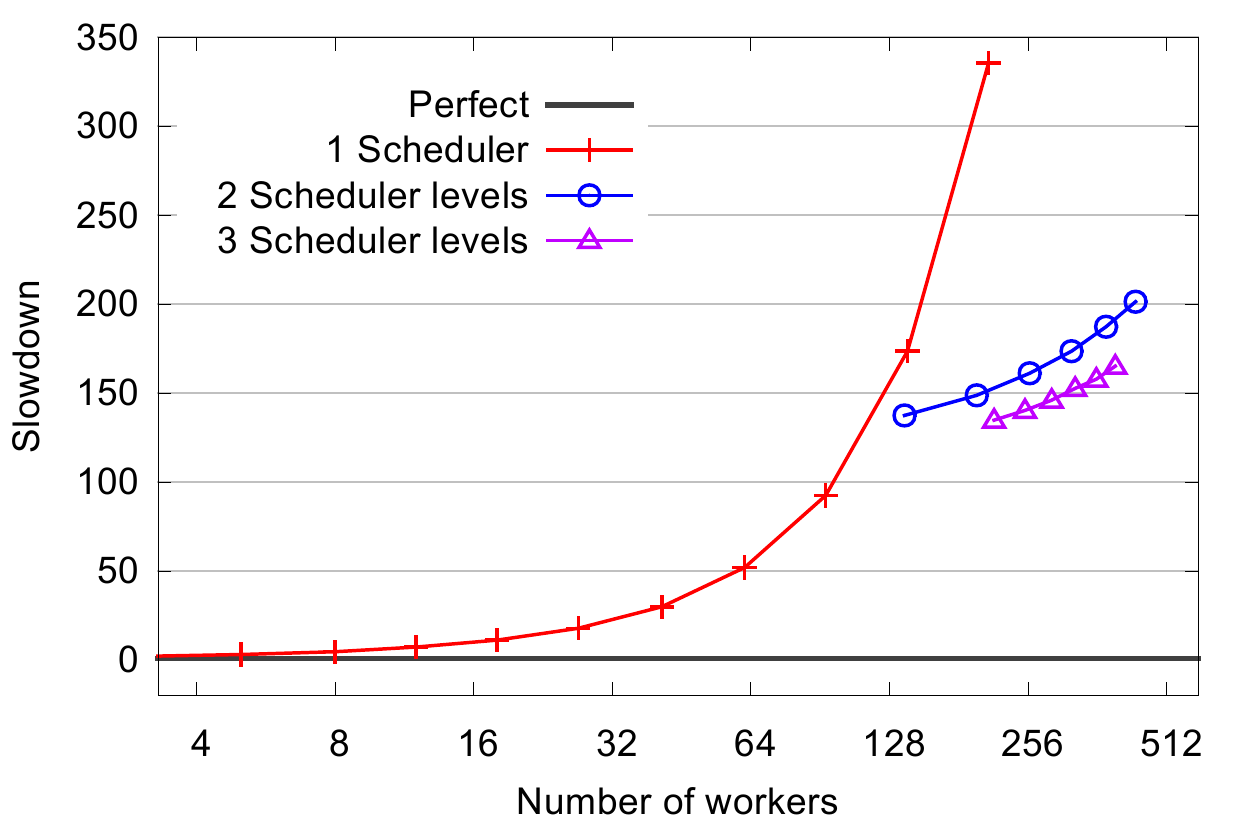}
    \label{fig:multilevel_weak}
  }
  \caption{
    Task granularity impact on a MicroBlaze scheduler (a).
    Multi-level benchmark weak scaling (b).
  }
\end{figure}

Our final experiments explore how Myrmics behaves using more
than two levels of schedulers. Since we are limited to eight ARM
Cortex-A9 cores, we use
only the 512-core MicroBlaze homogeneous system, where we can
use as many of its cores as we see fit to be schedulers. The
MicroBlaze-only system has different intrinsic overheads;
Fig.~\ref{fig:task_times} shows that the spawn
delay rises to 37.4 K cycles. To better understand how this affects
performance, we first repeat the task granularity
impact experiment using a single MicroBlaze scheduler, with the
same parameters as described in section~\ref{sec:overhead}.
Fig.~\ref{fig:task_synthetic_mb} shows the results for the
homogeneous system. Notice that the achievable speedup is much
lower for a single scheduler and that for low core counts, the optimal
number of workers for a given task size is given by dividing the task
size by 37.4 K.

In order to test for multiple scheduler levels, we saturate the
schedulers as much as possible. We use a synthetic benchmark that
creates a hierarchy of small regions and spawns empty tasks that do
nothing.  Each task normally takes 22.5 K clock cycles,
if the scheduler is under no load. Using the empty tasks, we
manage to saturate the schedulers so
that more than two levels of them are needed to satisfy all the
system load. Fig.~\ref{fig:multilevel_weak} shows the results of
this experiment. 
First, for a single scheduler (red/cross line) the slowdown
is excessive for high core counts, much more than
when the tasks have any significant size (such as the
Fig.~\ref{fig:task_synthetic_mb} behavior). 
Second, for the 2-level scheduling (blue/circle line) the hierarchy
of schedulers performs significantly better. This is consistent
with the behavior seen in all real benchmarks we
ran on the heterogeneous platform and present in
section~\ref{sec:scaling}. For the 2-level scheduling
experiments, we use a fanout of 6 for the scheduler-to-worker
ratio, \textit{i.e.} each low-level scheduler is responsible for
6 workers. We explored multiple alternatives and found that
this ratio is a good trade-off that makes the low-level
schedulers operate fast enough, without requiring too many
schedulers to be present\footnote{\ Recall that
the heterogeneous platform had an optimal
scheduler-to-worker ratio of 16.}. As the 2-level setup
scales, the top-level scheduler increasingly begins to saturate:
when we reach the point of 438 worker cores, there are 73 leaf
schedulers and 1 top-level scheduler. 
Third, using 3-levels of schedulers (purple/triangle line) reduces the
latency problems caused by the saturation of the top-level
scheduler and provides roughly a 15\% improvement on the
slowdown. We use again a scheduler-to-scheduler ratio of 6 for
the middle-to-leaf scheduler configuration. The
improvement is not so dramatic as the one from a single
scheduler to 2-level scheduling. Every additional level of
scheduling incurs an increase in overhead, as
it implies a more distributed region tree and inter-scheduler
communication to traverse it. Although it is a contrived
example, this experiment confirms that Myrmics can scale using
more than three levels of schedulers. We also validated the
system for correctness running benchmarks with four and five
levels of schedulers, which for the limited number of worker
cores exhibit a performance slowdown compared to the 3-level
setup.

\section{Related Work}
\label{sec:related}

\paragraph{Independent tasks and PGAS SPMD models}
The first task-parallel programming models that appeared assume
that any tasks spawned by the program can begin executing
immediately. The runtime system takes care of scheduling these
tasks, but does not automatically infer dependencies among tasks.
Typical examples of such programming models include
Cilk/Cilk++~\cite{cilk_implementation}, Intel Thread Building
Blocks (TBB)~\cite{tbb} and OpenMP
tasks~\cite{openmp_tasks}\footnote{\ OpenMP 4.0 supports expressing
explicit dependencies between tasks through the \texttt{depend}
directive.}, in which tasks are created
directly in the program or by a partitioning
algorithm~\cite{cilk_implementation,apcplus,tzannes:ppopp10}.
These systems focus on shared memory systems and focus on dataflow
dependencies among tasks; Myrmics targets a manycore architecture
without cache coherence where the runtime may also transfer task data.
Single-Program-Multiple-Data (SPMD)
programming models, such as Co-Array Fortran~\cite{caf},
UPC~\cite{upc,bupc}, and Titanium~\cite{titanium}, do not support
the tasking model but rely on automatically extracting parallel
behavior from a serial program. The languages provide the
Partitioned Global Address Space (PGAS), in which the user
specifies how arrays are distributed. System-wide memory accesses
(global) and thread-only accesses (local) are differentiated via
the type system. The
runtime system interferes to execute global memory accesses. Both
independent tasks and SPMD models rely on the user to write
hazard-free programs, as the runtime system can ensure
neither the correctness nor the determinism of the program.
Myrmics also implements a global address space and all tasks are
dependent on the memory accesses they make. The runtime system
can use this information to enforce correctness and determinism,
as well as exploit the data placement knowledge to optimize
scheduling for locality.  Such scheduling is difficult or impossible
using existing operating systems. Bubblesched~\cite{bubblesched} is a
user-level scheduler framework that accepts hints for various kinds of
optimizations, for cache coherent shared-memory systems.

\paragraph{Static dependency analysis}
An alternative approach to task parallelism is to rely on the compiler
to perform static analysis, in order to discover which tasks can
be safely run in parallel. Static analysis techniques can be
quite complex, lead to imprecise results and may require
significant compilation time, depending on the application. When
successful, they can offer correctness guarantees to the user and
they alleviate the runtime system from a lot of overhead, as it
can disregard dependency checks from many spawned tasks. Dynamic
Out-of-Order Java~\cite{doj} is a recent example of such a
language, which performs many static optimizations and also uses
\textit{heap examiners} at runtime, to resolve cases that are
ambiguous by the static analysis. Myrmics does not rely on static
analysis techniques, but is still able to use compiler hints to
exclude certain task arguments if the compiler marks them as
\texttt{SAFE} ---our modified SCOOP compiler~\cite{scoop_pact12}
can provide Myrmics such hints.

\paragraph{Dynamic dependency analysis}
The newest class of task-parallel programming models enables the
user to write serial code split into tasks, which are not
necessarily allowed to run immediately when spawned. The user
specifies constraints to inform the runtime system when a task
should be allowed to run.  Such tasks are similar to
codelets~\cite{codelets,runnemede}, namely tasks with clearly defined
inputs and outputs that run to completion without interrupt.
The OmpSs family of programming
models~\cite{starss,ompss} lets the user annotate serial
code with compiler pragmas that inform the runtime which
variables will be touched by each task and how (read or written).
A source-to-source compiler translates the pragmas into runtime
hooks, which call the runtime library to perform dynamic task
dependency analysis.  Myrmics follows a similar approach. In
contrast to Myrmics, the OmpSs models support expressive formats
for array portions, such as strides or dimension parts, but do
not support pointer-based data structures. Myrmics implements a
global address space with software-guaranteed coherency, which
resembles ClusterSs, the OmpSs variant for
clusters~\cite{cluster_ss_2,cluster_ss_1}. However, ClusterSs has
a centralized directory to track object locations across the
cluster which inherently limits its scalability.
The StarPU~\cite{starpu} and XKaapi~\cite{xkaapi} runtimes implement
similar task-parallel dataflow programming models for heterogeneous
CPU/GPU shared memory architectures.  As in Myrmics, they hide
underlying data transfers from the programmer.  These systems,
however, use a centralized scheduler to solve dependencies and
schedule tasks on CPU threads and GPUs, which performs well for low
numbers of cores.

The above task-parallel programming models restrict task inputs and
outputs to objects, memory ranges or tiles, requiring additional
programmer effort and encodings to express tasks operating on dynamic
data structures like lists or graphs; in comparison, Myrmics uses the
region abstraction to better express such tasks.
Myrmics support for regions also resembles the Legion programming
language~\cite{legion,legion_tr}, where the user can specify
logical collections of objects and their mapping onto the
hardware. Myrmics has a different focus than Legion, as it
explores at depth the implementation aspects on how
dependent-task runtime systems can be structured to scale on
manycore architectures.  The authors of Legion focus instead on
the language structure and do not supply enough runtime system
details on how they distribute the system load, or how well their
implementation scales to high core counts. We specifically create
and benchmark Myrmics on a 520-core single-chip architecture to
explore scalability problems.

Legion is a generalization of Sequoia~\cite{sequoia}, which introduces
hierarchical memory
concepts and tasks that can exploit them for portability and
locality awareness. X10~\cite{x10} also supports regions, but
these refer to parts of multi-dimensional arrays and can be
extracted by the compiler. ParalleX~\cite{parallex} is a an
asynchronous parallel computing model with a partitioned global
address space that evolved from EARTH~\cite{earth} dataflow model.
ParalleX combines high-level parallelism and synchronization
abstractions such as asynchronous calls, futures and atomic sections
with partitioned address space and message passing.
Chapel~\cite{chapel,chapel2}
introduces the \textit{domain} concept to support
multi-dimensional and hierarchical mapping of indexes to hardware
locations. Another programming model that uses regions is
Deterministic Parallel Java (DPJ)~\cite{dpj}, which is a parallel
extension of Java. DPJ combines compiler optimizations and
dynamic runtime checks to guarantee determinism and to maximize
available parallelism. Like Myrmics, DPJ supports hierarchical
regions. Another way to spawn dependent tasks is to use
\textit{futures}, which declare that a new task must wait for
certain variables (Data-Driven Tasks~\cite{data_driven_tasks} and
X10), or other tasks (Habanero-Java~\cite{habanero}). Finally,
OpenStream~\cite{openstream} offers another alternative to
enhance the OpenMP tasking model to support data-flow
parallelism, through the use of \textit{streams}, which defines
how the tasks produce and consume data.

%
%
%

\section{Discussion}
\label{sec:discussion}

Throughout the design, development and evaluation of the Myrmics
runtime system, we delved into various scalability problems.
In this section we discuss some key issues and share our
insight on how these may stimulate further work in this area.
First, we observe that the dominating factor affecting the performance
of any task-parallel runtime system is the duration of the spawned
tasks, \textit{i.e.} the task granularity; bigger tasks reduce the
perceived per task overhead.
Picking the ``correct'' task size depends on a lot of factors, such as
how much data a task touches and how these data fit into the cache
and memory hierarchy.  Smaller task sizes expose more parallelism and
exhibit better cache hit ratios.
Future runtime systems need to scale to hundreds of cores, but if this
is done simply by increasing the task size to minimize the runtime
overhead, it will hurt parallelism. To make the most out of the
emerging manycore architectures, we need 
\textit{scalable} solutions for scheduling and dependency analysis
algorithms.

A second factor is the number of ready tasks: Assuming that the
runtime overhead and problem size allows it, is it worth
over-decomposing a problem to very small tasks? For a given per-task
overhead, more smaller tasks will incur more total overhead. On the
other hand, when worker queues are kept non-empty, the runtime can
prepare the next task by transferring its data during the execution of
the current task (as in Myrmics) or by enabling a hardware-assisted
prefetcher to touch the needed cache lines and bring them closer (in a
cache-coherent system).  Based on our experience with most benchmarks,
we conjecture that having ready tasks that are twice the number of
cores is a good trade-off point for decomposing the problem.

Third, 
we estimate that in future manycore chips CPU cores must diversify
their functionality. Our experience with two core roles (scheduler and
worker) in Myrmics is very positive, as it allows uninterrupted
execution of worker tasks and fast processing of scheduler events.
Although we do not yet have an alternative implementation to measure
against, we believe that core specialization will lead to improved
cache hit ratios and avoid context switching overhead.
Organizing the cores in a strict hierarchy has proven to be
advantageous, yet cumbersome. 
Our insight is that the hierarchical
organization successfully manages to localize application parts around
the low-level schedulers. The communication and data movement remain
localized for much of the application running time and can thus
decrease network traffic and increase the application and power
efficiency. However, restricting the communication only across
parent-children cores tends to be also problematic in handling a few
corner cases.
Our measurements reveal that increasing the levels of
the hierarchy seems promising to scale the runtime system to many
hundreds of cores, but this does come with an added overhead per
scheduler level. Tuning the runtime system to a ``correct''
balance of schedulers and workers is a trade-off.  More schedulers
balance the runtime load and enable faster event processing, but also
increase the average per-task overhead, especially for non-leaf tasks.

\section{Conclusions}
\label{sec:concl}


This work explores some of the challenges that lie ahead for
task-parallel runtime systems running on emerging processors with
hundreds of CPU cores. We present the Myrmics runtime system and
evaluate it on a prototype platform that emulates a single-chip
processor of 8 strong cores and 512 weaker cores. Scaling a
runtime system to hundreds of cores using fine-grain tasks is
very challenging. We explore several concepts towards this goal,
such as specializing CPU roles, hierarchical organization,
limiting task argument expressiveness and selecting an optimal
task granularity. Our measurements suggest that our main
hypotheses are supported and that many of these ideas are
promising. Current hardware architecture trends hint that runtime
systems must evolve fast to catch up with the increasing number
of cores; radical changes may be needed, especially if the new
processors become more heterogeneous and/or less cache coherent
than today's norm. We think that our work poses interesting
questions on enhancing runtime system scalability and that it
will stimulate further research.

\bibliographystyle{IEEEtran}
\bibliography{references}

\end{document}